\def\anc{{$d$-Al-Ni-Co}\/}
\def\alpm{{$i$-Al-Pd-Mn}\/}

\documentclass[prb,preprint,amsmath,amssymb,aps,floatfix,epdf,psfig]{revtex4-2}
\usepackage{epsf}
\usepackage{amssymb}
\usepackage{amsmath}
\usepackage{graphicx}
\usepackage{verbatim}
\usepackage{dcolumn}
\usepackage{bm}
\usepackage{multirow}
\usepackage{array}
\usepackage{booktabs}
\usepackage{amsmath}
\usepackage[normalem]{ulem}
\usepackage{xr}
\usepackage{xcolor}
\usepackage{txfonts} 
\usepackage{pgfmath}
\usepackage{pgffor}
\usepackage{dsfont}
\usepackage{bm}
\usepackage[cp1250]{inputenc}
\usepackage{color}
\usepackage{colortbl}
\usepackage{multirow}
\usepackage{wrapfig}
\usepackage{wasysym}
\usepackage{physics}
\usepackage{array}

\usepackage[hypertexnames=false,linktocpage=true]{hyperref}
\hypersetup{colorlinks=true,linkcolor=blue,anchorcolor=blue,citecolor=blue,menucolor=blue,filecolor=blue,urlcolor=blue}

\pdfoutput=1 

\begin{document}
\title {\textbf{Decagonal  Sn clathrate on $d$-Al-Ni-Co}}	
	\author{Vipin Kumar Singh$^{1}$, Eva Pospisilova$^{2}$, Marek Mihalkovi\v{c}$^{2\dagger}$,
		Marian Kraj\v{c}\'i$^{2*}$, Pramod Bhakuni$^{1}$,  Shuvam Sarkar$^{1}$,  
		Katariina Pussi$^{3}$,   D. L. Schlagel$^{4}$, T. A. Lograsso$^{4}$,   Paul C. Canfield$^{4,5}$, Sudipta Roy Barman$^{1\#}$} 
	
	\affiliation{$^{1}$UGC-DAE Consortium for Scientific Research, Khandwa Road, Indore - 452001, Madhya Pradesh, India}
	\affiliation{$^{2}$Institute of Physics, Slovak Academy of Sciences, D\'ubravsk\'a  cesta 9, SK-84511 Bratislava, Slovak Republic}
	\affiliation{$^{3}$LUT School of Engineering Science, P.O. Box 20, FIN-53851 Lappeenranta, Finland}
	\affiliation{$^{4}$Ames Laboratory, Iowa State University, Ames, Iowa 50011, USA}
\affiliation{$^{5}$Department of Physics and Astronomy, Iowa State University, Ames Iowa 50011, USA	}

		\begin{abstract}
Decagonal quasiperiodic ordering of Sn thin film on \anc\,  is shown  based on  scanning tunneling microscopy (STM),   low-energy electron diffraction and density functional theory (DFT).
~Interestingly, the decagonal structural correlations are partially retained up to a large film thickness of 10 nm grown  at a  165$\pm$10~K. The nucleation centers called ``Sn white flowers''  identified by STM at submonolayer thickness  are recognized as valid patches of the decagonal clathrate structure with low adsorption energies of the SnWF motifs. Due to the excellent lattice matching (to within 1\%) between columns of Sn dodecahedra in the clathrate structure and pentagonal motifs at the \anc\ surface, the interfacial energy favors clathrate over the competing Sn crystalline forms. DFT study of the Sn/Al-Ni-Co composite model shows good mechanical stability, as shown by the work of separation of Sn from Al-Ni-Co slab that is comparable to clathrate self-separation energy.  The relaxed surface terminations of the R$_2$T$_4$ clathrate approximant are in  self-similarity correspondence with the motifs observed in the STM images from monolayer to thickest Sn film. 
\end{abstract}
\maketitle

\section{Introduction} 
		Quasicrystals show sharp diffraction spots although they do not  have translational long range order and rather possess forbidden rotational symmetries. In addition to intermetallic alloys~\cite{Shechtman_prl84,Tsai_nat00}, quasicrystallinity has also been observed in various forms such as  colloidal systems~\cite{Fischer_pnas11}, binary nanoparticles super lattices~\cite{Talapin_nat09},  molecular	assemblies~\cite{Xiao_nat12,Ye_Natmat17},  chalcogenides~\cite{Cain_pnas20}, twisted bilayer graphene~\cite{Ahn_sci18,Yao_pnas18} and even in 	naturally occurring minerals~\cite{Bindi_pnas12}.  Quasicrystals are fascinating, especially due to their unusual physical properties such as low specific heat, low thermal and electrical conductivities, and low friction
		~\cite{Kang_jmr93, Dubois_mse94, Poon_ap92}; as well as evidence of Anderson localization~\cite{Sarkar21}, demonstration of pseudogap at the Fermi level~\cite{Nayak12}, and recent  prediction of topological states~\cite{Chen_prl20, Fan_fp22}.  

The basic entities that are known to form quasicrystals are clusters such as pseudo-Mackay and Bergman clusters~\cite{Guyot_crp14}.  Crystalline structures with clusters entities, for instance, clathrates, somewhat resemble the clusters that form the quasicrystals. The most straightforward rationalization of the bulk clathrate structure is via the network of its cage centres~\cite{Frank_58, Frank_59}.  Polytetrahedral order~\cite{Nelson_ssp89,Sadoc_99} is the basis of the crystalline Frank-Kasper phases~\cite{Frank_58,Frank_59,Shoemaker_88} and has been used in order to realize the structure of quasicrystals~\cite{Elser_prl89,Audier_pm86,Sachdev_prb85}. Materials exhibiting clathrate structure~\cite{Schafer_jacs13,Sun_prl19,Lin_sci18,Zhu_Sciadv20,Ikeda_ncom19,Zhu_prl20,Ghosh_pnas19} have become a subject of interest  due to their optical and thermoelectric properties~\cite{Iversen_jssc00,Takabatake_rmp14}, as well as due to their potential for application~\cite{Kume_jjap17,Anna_mse16}. Guest-free Si~\cite{Gryko_prb00} and Ge~\cite{Guloy_nat06} clathrates have wide quasi-direct band gaps that is important for photovoltaic applications, and Sn  with guest atoms has shown clathrate structures~\cite{Christensen_jap13}.

 Since the last decade, Sn films have attracted considerable attention because of their interesting  electronic properties~\cite{Yuhara_2dmat18, Sadhukhan_prb19, Sadhukhan_ass20, Rogalev_prb19, Liao_natphy18, Sante_prb19}. In particular,  in its single layer honeycomb structure (stanene), Sn has been shown to exhibit  topological properties~\cite{Shi_pla20, Si_jpcl20,Xu_13}. In a recent advancement towards achieving elemental quasicrystal,  we have shown  using scanning tunneling microscopy (STM), low energy electron diffraction (LEED), and density functional theory (DFT)  that Sn can grow quasiperiodically  up to about 4 nm thickness with a clathrate structure on  $i$-Al-Pd-Mn substrate~\cite{Singh_prr20}.  The clathrate quasiperiodic structure nucleates on \alpm\, because of  an excellent  matching (within 1\%) of  the  cage-cage linkage length in Sn clathrates  ($\approx$1.26 nm) with  the pseudo-Mackay cluster-cluster separations in $i$-Al-Pd-Mn (1.255 nm).

In the present study, we explore the growth of Sn films on  decagonal ($d$)-Al-Ni-Co quasicrystal with a motivation  to demonstrate the generality of  the clathrate quasiperiodic structure of Sn. \anc\, has a different structure compared to \alpm: the former comprises of decagonal  quasiperiodic planes, but  exhibits  translational periodicity  along the ten fold (10f) axis with an inter-planar distance of approximately 0.2 nm~\cite{Tsai_mattra89_2}. The structure of \anc\, has not been solved experimentally until date, but several models have been proposed in literature~\cite{Hiraga94,Yamamoto97,Sugiyama02}. Sugiyama {\it et al.}  determined the  structure of  the W approximant of \anc\, (W-Al-Ni-Co)  using  x-ray diffraction, and showed that it is  closely related to \anc~\cite{Sugiyama02}. The 10f \anc\, surface  was modelled by Kraj\v{c}\'{i} \textit{et al.}~\cite{Krajci06} based on W-Al-Ni-Co as alternating flat (\textit{A}) and puckered (\textit{B}) atomic layers perpendicular to the periodic  axis along the $z$ direction. The surface energy of $d$-Al-Ni-Co is reported to be  1.17 J/m$^2$~\cite{Fournee_ss03}, while that of Sn is 0.71 J/m$^2$~\cite{Tyson_ss77}. Thus, from surface energy considerations, Sn is expected to wet the \anc\, surface. However, Sn deposition on \anc\, has been scarcely studied in literature.  Shimoda $et~al.$ reports a pseudomorphic monolayer of quasiperiodic Sn on $d$-Al-Ni-Co using RHEED, XPS, and STM~\cite{Shimoda_jns04}. For depositing Sn, these authors however pre-coated the backside of the specimen with  Sn and then with heating Sn diffused and spread over to the front side. This method precludes  possibility of growing thicker Sn films, since it requires high temperatures. 

Although there has been four decades of active research on quasicrystals,  elemental quasicrystals  have remained elusive so far. In this paper, we show the formation of  decagonal Sn thin film (thickness $<$1 nm) on a 10f $d$-Al-Ni-Co surface   based on LEED and STM. The decagonal structural correlations are partially retained up to a large film thickness of 10 nm. The quasiperiodic motifs observed from STM  up to 10 nm thick Sn film are in good self-similarity correspondance  with the decagonal clathrate  R$_2$T$_4$ approximant model of Sn calculated by DFT  that show relaxed penta-hole  and penta-cap terminations.   The ``Sn white flowers'' (SnWFs) identified by STM after submonolayer Sn deposition are the nucleation centers of the decagonal clathrate structure with low adsorption energies. Due to the excellent lattice matching (to within 1\%) between Sn  clathrate and the \anc\ surface, the interfacial energy favors the clathrate. DFT study of the Sn/Al-Ni-Co composite model shows good mechanical stability, as shown by the work of separation of Sn from Al-Ni-Co slab that is comparable to clathrate self-separation energy.  The paper is organized as follows:  We provide experimental evidence of  decagonal quasicrystallinity in  Sn thin films  using STM and LEED (subsection \ref{subsec:monolayerSn}) and the evidence of decagonal structural correlations in  the thick films is provided in  subsection \ref{subsec:thickSnfilm}. Thereafter, we provide a theoretical model for a decagonal clathrate structure of Sn from DFT  (subsection \ref{subsec:surfmodel}). A piece of evidence supporting the clathrate decagonal model is a detailed analysis of the nucleation supported by the STM results that are discussed   in subsection~\ref{section:nucleation}.  
~Next, in subsection \ref{subsec:Snonsubstrate}, we consider a Sn/$d$-Al-Ni-Co R$_2$T$_4$ approximant composite model to study the adlayer-substrate interaction with DFT. Finally, in  subsection \ref{subsec:theoryandstm} we define the  motifs based on our theoretical model and show that  these are observed  from STM  and a sizable contagious region of the image for the 10 nm thick film is in good agreement with the theory.  
 
\section{Methods}    
The STM experiments were carried out in a variable temperature STM system from Scientaomicron at a base pressure of 5$\times$10$^{-11}$mbar. The experiments were performed  using an electrochemically etched polycrystalline tungsten tip that was  cleaned \textit{in situ} by Ar$^{+}$ ion sputtering and voltage pulse method. The images were recorded in the constant current mode with the sample at ground potential for various sample bias voltages and are shown after low-pass Fourier transform filtering. The  analysis was performed  using the
SPIP software.   The zero of the color scale for all the STM images  corresponds to the bearing height \textit{i.e.} the most frequently occurring  height. The  root mean square roughness  ($S_q$) is defined as  the square root of the sum of squares of each height value ($z$) at each pixel coordinate ($x_k$, $y_l$) in the ($M$$\times$$N$) pixel dataset given by 	$S_q=\sqrt{\frac{1}{MN}\sum_{k=0}^{M-1}\sum_{l=0}^{N-1}[z(x_k,y_l)]^2}$.
 
 The LEED equipment   from OCI Vacuum Microengineering has a retractable  four grid rear view optics, and the patterns were recorded with a digital camera in nearly normal incidence geometry at 1 eV step of the beam energy ($E_p$) that was varied from 30 to 200 eV. The I-V curves were determined for a spot by calculating its intensity for each $E_p$ in a window of fixed size with the spot in its middle. An averaging was performed for all the symmetry equivalent spots.   Image J software~\cite{Schneider_natmeth12}  was used to invert the gray color scale after applying the auto contrast option  to adjust the brightness and intensity. 

Monocrystalline $d$-Al-Ni-Co quasicrystal was grown from a high temperature ternary melt~\cite{Canfield1}  and separated from the excess liquid via centrifuging~\cite{Canfield2}.
	~\anc\, was mounted on a  specially fabricated cooling sample plate made of molybdenum. A smooth and  clean  $d$-Al-Ni-Co surface was obtained by cycles of 0.5-1.5 keV Ar$^{+}$ ion sputtering and followed by annealing at 1043 K in the UHV chamber for 4 hr~\cite{Singh_aip20}.  Sn of 99.99\% purity was evaporated using a water-cooled Knudsen evaporation cell~\cite{Shukla_rsi04} equipped with a shutter that is operated manually using a rotary feedthrough. The temperature of the cell used for deposition was 1163 K and this was measured by a K-type thermocouple placed at the outside bottom of the pyrolytic boron nitride crucible. The deposition was performed  at an  angle of 70$^\circ$ from the surface normal, and the pressure of the chamber during the deposition was 5$\times$10$^{-10}$ mbar.  The substrate surface was freshly prepared for both  0.2 ML and 1 ML  Sn deposition. 
		~For the thick films with $t_d$ varying from 3 to 37 min, the deposition was sequential. For the depositions at 165$\pm$10~K, the LEED was performed at the same temperature while  STM was performed at 80~K.   
		\begin{figure}[htb]
	\centering
	\includegraphics[width=160mm,keepaspectratio]{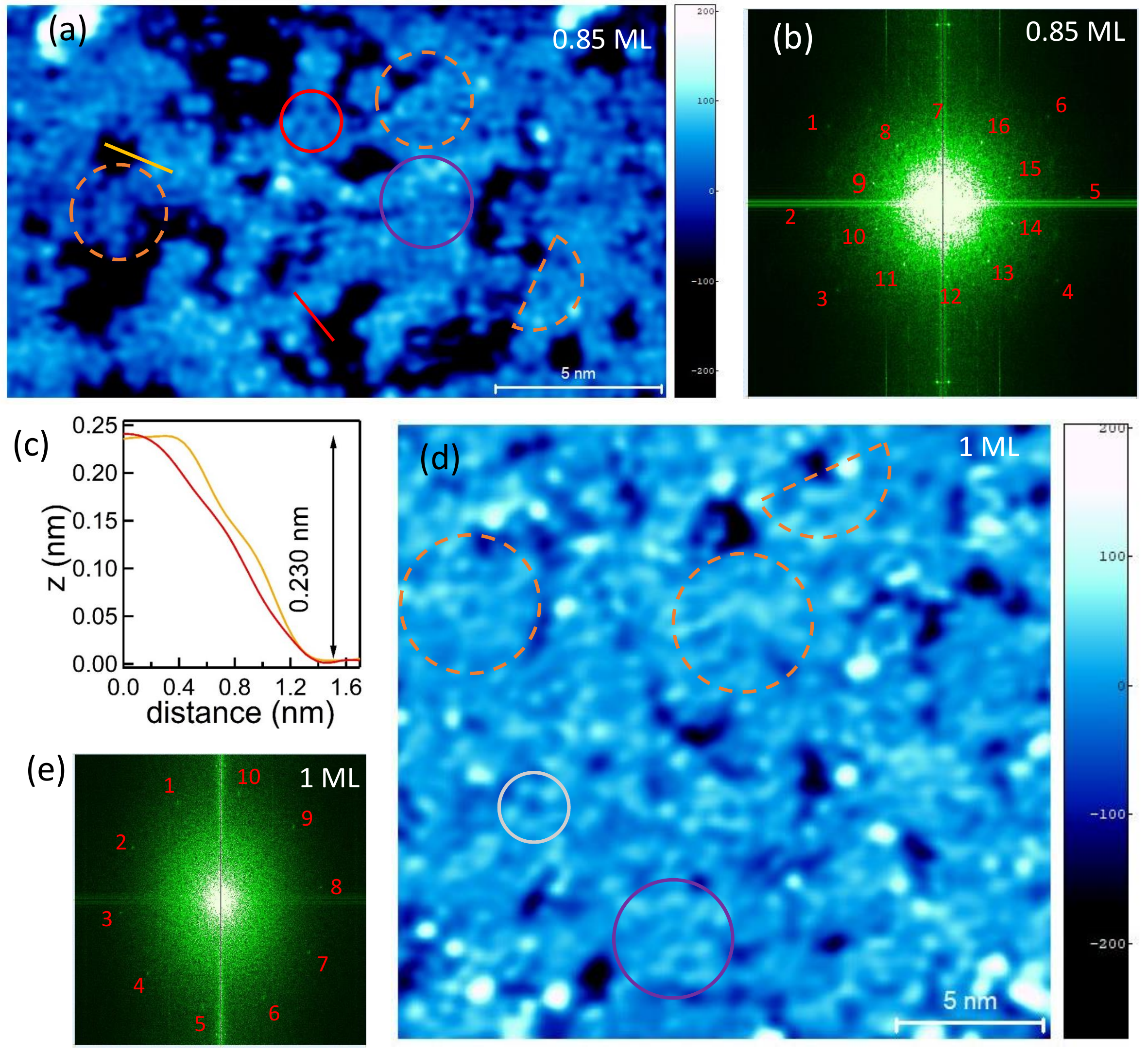} 
	\caption{(a)  STM topography image of 0.85 ML Sn/$d$-Al-Ni-Co ($I_T$= 0.5 nA, $U_T$= -1.5 V) at 300~K. 	The color scale representing the height in picometer is shown on the right, zero corresponds to the bearing height.  (b) Fast Fourier transform (FFT) of panel~\textbf{a}, the two sets of decagonal spots are numbered.  (c) The height profile along the red and yellow lines  in panel \textbf{a}, the double arrow corresponds to the thickness of the Sn monolayer islands. (d)   STM topography image ($I_T$= 0.6 nA, $U_T$= -1.5 V) of 1 ML Sn/$d$-Al-Ni-Co. In panels \textbf{a} and \textbf{d}, the different motifs such as wheel  (dashed orange circle),  pentagon (gray circle),  polygon assembly (violet circle), and crown (dashed orange half-circle) are highlighted.  See subsection \ref{subsec:theoryandstm} and   Fig.~\ref{motifs_theory} for their description, same line-type 
		~is used henceforth for  highlighting the each type of motif.   (e) FFT of  panel \textbf{d} with the decagonal spots numbered 1-10.} 
	\label{1ML_STM_RT}
\end{figure}

Structural optimization study of the Sn surface terminations and reconstructions, as well as Sn/Al-Ni-Co model relaxations, have been performed using plane-wave DFT code Vienna {\em ab initio} Simulation Package (VASP) ~\cite{Kresse_prb96}.  We used projector augmented wave potentials~\cite{Kresse_prb99} in the PW91 generalized gradient approximation~\cite{Perdew_prb92}. Default energy cutoff ($ENCUT$) 103.3 eV preset by ``accurate'' calculational setup turned out to be sufficient considering {\em energy differences} $\Delta E$ between different samples: out of the four samples examined for convergence, change of the $\Delta E$ from $ENCUT$= 103.3 to 300 eV was $\leq$0.3 meV/atom, certainly well below other unaccounted systematic inaccuracies.  The $k$-point meshes were converged to comparable change in $\Delta E$ upon increasing mesh density.

\section{Results and Discussion}
	\subsection{Quasiperiodicity of Sn thin films  ($<$~1~nm) on \anc}
	\label{subsec:monolayerSn}

\subsubsection{\underline {Monolayer Sn:}}
	
 An  STM topography image  of 0.85 ML Sn  on $d$-Al-Ni-Co 
 ~ in  Fig.~\ref{1ML_STM_RT}(a) indicates possible quasiperiodicity of the adlayer through  sharp 10f spots in its fast Fourier transform (FFT) in Fig.~\ref{1ML_STM_RT}(b) that has \AC85\% contribution from the adlayer. The FFT  shows  two sets of 10f spots. The spots are numbered such that the corresponding intensity profiles  along the tangential direction can be shown.  In Fig.~S1(a,b) of the supplemental material (SM)~\cite{Supp},  each intensity profile shows a peak at the position of the spot and thus all the spots are unambiguously identified.  The ratio of their radii 
 ~  is  1.92$\pm$0.03, which is  close to $\tau\chi$= 1.90, where $\tau$ is the golden mean given by $\tau$= (1+$\sqrt{5}$)/2= 1.618. $\chi$  is the ratio of the side of  a regular pentagon  and the distance from its center to the vertex; it is related to $\tau$ by   $\chi$ = $\sqrt{(3-\tau)}$= 1.176~\cite{Sharma_prb04,Haibach1999}. 
  The outer set of spots is  rotated by 18$^{\circ}$ with respect to the inner set. The height profiles taken along the red and yellow lines from the STM image  [Fig.~\ref{1ML_STM_RT}(a)] are plotted in Fig.~\ref{1ML_STM_RT}(c).  These show  the difference of the average $z$ corrugation (the double-sided arrow) of  the  Sn monolayer islands (light blue regions) and the substrate (dark), which is the thickness of the latter under the assumption that the electronic height is equal to the geometric height. After averaging over more than 30  such profiles from different parts of the image as in our earlier work~\cite{Rai2019}, we find the thickness of the Sn monolayer island  to be 0.23$\pm$0.02 nm.  
 

 In Fig.~\ref{1ML_STM_RT}(d),  STM topography image of  nearly one monolayer (0.97 ML) is displayed.  The 10f symmetry of the spots  observed in the FFT [Fig.~\ref{1ML_STM_RT}(e)] and their intensity profiles  along the tangential direction are shown in Fig.~S1(c) of SM~\cite{Supp}.  Two adjacent spots subtend  an angle of 36$^{\circ}$$\pm$2$^{\circ}$  at the center demonstrating the quasiperiodic nature of the Sn monolayer. Different quasiperiodic motifs of Sn such as wheel, crown,  pentagon, and polygon assembly are highlighted in Figs.~\ref{1ML_STM_RT}(a,d). These motifs are defined on the basis of the relaxed surface of the clathrate structural model calculated by DFT (see subsection \ref{subsec:theoryandstm}).

			\begin{figure}[t]
				\centering
				\includegraphics[width=170mm,keepaspectratio]{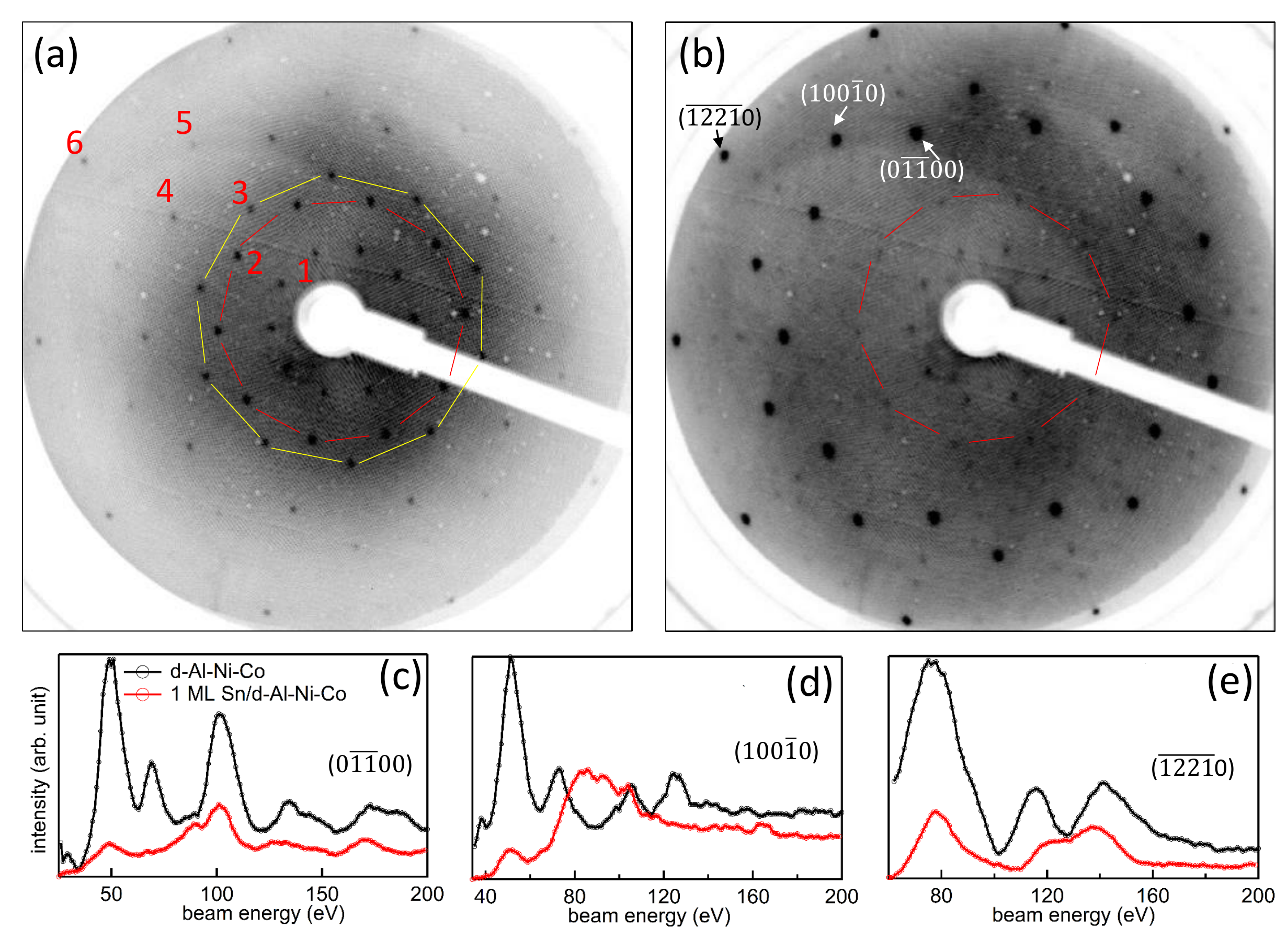}
				\caption{The LEED pattern  of (a) 1 ML Sn/$d$-Al-Ni-Co deposited at  165$\pm$10K (LT) is  compared to   (b) the substrate $d$-Al-Ni-Co. Both the patterns are shown in an inverted gray scale and are taken with  beam energy ($E_p$) of  55 eV. The different sets of decagonal spots are numbered 1-6 in panel \textbf{a}.  The red and yellow lines highlight the 2$^{\rm nd}$ and the 3$^{\rm rd}$ set, respectively (the latter is not visible in panel \textbf{b}).  Comparison of the  I-V curves in the range 30$\leq$$E_p$$\leq$200 eV for (c) (0$\bar{1}\bar{1}$00), (d) (100$\bar{1}$0) and (e) ($\bar{1}\bar{2}\bar{2}\bar{1}$0) spots between 1 ML Sn and the substrate.}
				\label{1ML_LEED_LT}
			\end{figure}

We find that the growth of  Sn films of larger thickness beyond a monolayer is difficult at 300~K because of  very low sticking coefficient and large diffusivity. On the other hand, at liquid helium temperatures, the diffusivity would be too low for an ordered growth.  So, we have used an intermediate lower temperature (LT)   of 165$\pm$10K for growing  Sn films thicker than a monolayer.   LT is in fact  a stringent requirement for growing Sn thick films.

The quasiperiodicity of the Sn monolayer at LT  is  evident from several sets of 10f spots  in the LEED pattern (numbered 1-6)  at an angular separation of 36$^{\circ}$  [Fig.~\ref{1ML_LEED_LT}(a)]. The ratios of the radii of  the outer sets and the innermost set are  related by powers and product of $\tau$ and $\chi$, as shown in Table~SI of SM~\cite{Supp}. The intensity profile through some of the spots (numbered 1-10) of both the substrate and the Sn adlayer is shown in Fig.~S2 of SM~\cite{Supp}. To be noted is that the position of the spots change with $E_p$ (Fig.~S3 of SM~\cite{Supp}). In particular, for $E_p$= 100 eV, the prominent intensities are seen at a higher reciprocal distance compared to the substrate in contrast to $E_p$= 55 eV (Fig.~\ref{1ML_LEED_LT}). Note that the LEED pattern is  significantly different from  the substrate: \textit{e.g.} in Fig.~\ref{1ML_LEED_LT}(a), an extra set of diffraction spots (highlighted by yellow lines) is observed, which is not visible in  \anc\,  in Fig.~\ref{1ML_LEED_LT}(b).    Also, the  set of spots highlighted by red lines is more intense compared to  the substrate. The shape  of the intensity versus voltage (I-V) curves for the LEED spots as a function of $E_p$  depends sensitively on  the surface structure~\cite{Heinz_95,Diehl_jpcm04,Pussi_prb06,Sarkar_apss21}. A comparison of  the I-V curves of monolayer Sn  in Figs.~\ref{1ML_LEED_LT}(c-e) for three different set of spots shows that these are quite different from the  substrate.  For example,  for the  (0$\bar{1}\bar{1}$00) spot, the substrate peak at 73 eV is completely absent, while the most intense peak at 52 eV is largely suppressed and shifted below 50 eV. It is a similar situation for the    (100$\bar{1}$0) spot below 80 eV, and furthermore the 126 eV substrate peak is absent. For the  ($\bar{1}\bar{2}\bar{2}\bar{1}$0) spot, the two peaks of the substrate at 116 and 141 eV move closer in the Sn layer and fill up the dip at 128 eV. The above discussed differences in the LEED evident from the patterns as well as the I-V curves 
 ~indicate   that although the Sn monolayer exhibits decagonal symmetry, it has a different structure portraying a non-pseudomorphic growth.
 
 A comparison of  the STM topography image of  the  Sn monolayer  at LT (Fig.~S4(a) of SM~\cite{Supp}) 
 with that at 300~K [Fig.~\ref{1ML_STM_RT}(d)]  shows that uniform wetting occurs at LT and  their roughness is almost similar  ($S_q$= 0.053$\pm$0.004 nm at LT, 
 ~whereas  it is 0.046$\pm$0.004 nm at 300~K). Motifs such as wheel, crown and  polygon assembly are also observed [Figs. S4(b-d)~\cite{Supp}]. 

	\begin{figure}[!htb]
	\includegraphics[width=170mm,keepaspectratio]{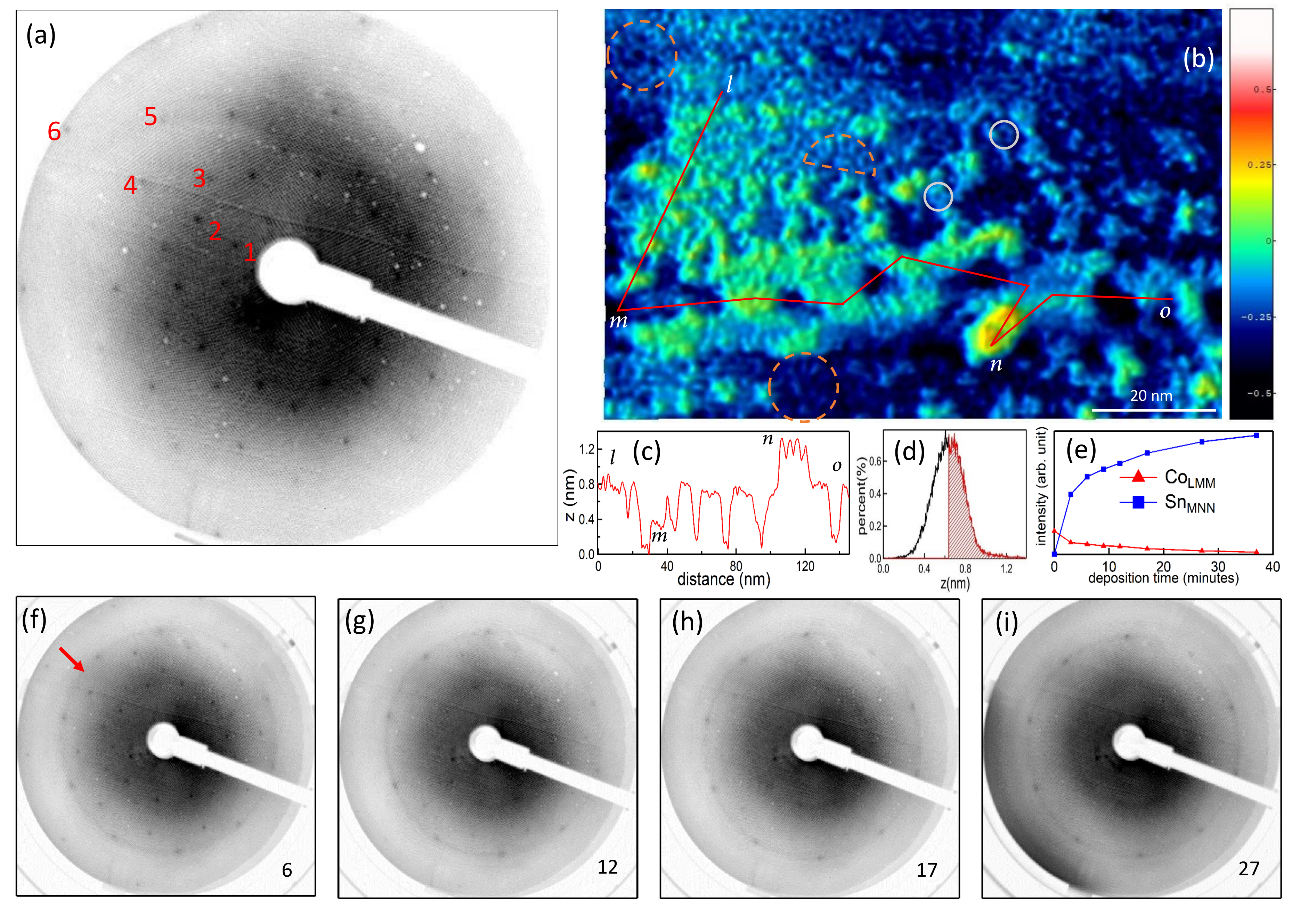}
	\caption{(a)  LEED pattern  of a Sn thin film (0.9 nm thickness) deposited at LT  on $d$-Al-Ni-Co. 
		~The pattern is shown in an inverted gray scale and is taken with $E_p$= 55 eV,  the different sets of decagonal spots are numbered as 1-6. (b) An STM topography image
		~for the same deposition with $I_T$= 1.5 nA, $U_T$= 1.5 V. 	 (c) The height profile along the line \textit{lmno} shown by red color in panel \textbf{b}.   (d) The height histogram of the panel \textbf{b}.  (e) Intensity variation of the Sn \textit{MNN}  
	~ and  the Co \textit{LMM} Auger signals as a function of $t_d$. 
		~  (f-i) The LEED patterns Sn films deposited for 6 min$\leq$ $t_d$ $\leq$ 27 min, as shown at the lower right corner of each panel ($E_P$= 55 eV).} 
\label{leed_stm_intrm}
\end{figure}

\subsubsection{\underline {Sn film of 0.9 nm  thickness:}}

	In Fig.~\ref{leed_stm_intrm}(a), the $t_d$=   3 min  LEED pattern shows six sets of 10f spots at angular separation of about 36$^{\circ}$ that are similar to the Sn monolayer. In Fig.~S5(a) of SM~\cite{Supp}, the spots are numbered  and the corresponding intensity  profiles are shown in Fig.~S5(b-e)~\cite{Supp}.  A video file named as  ``0.9nm"   in the SM~\cite{Supp} shows that the  intensities of all the spots in each set remain similar for the whole range of $E_p$ \textit{i.e.} 20$\leq$$E_p$$\leq$150 eV.  The above observations establish that this film exhibits  decagonal  symmetry.  For comparison, the LEED video corresponding to the monolayer is also uploaded as   ``monolayer" in the SM~\cite{Supp}.  The ratios of the radii of  the outer sets and the innermost set are related by powers and product of $\tau$ and $\chi$, as shown in Table~SII of SM~\cite{Supp}.
	
	 A STM topography image of this film in Fig.~\ref{leed_stm_intrm}(b) shows an increase in roughness with $S_q$ being 0.15 nm. Motifs highlighted by dashed orange and gray circles exhibit resemblance with the monolayer. The crown and the wheel motifs   [zoomed and compared with our theoretical model in Figs.~S6(a,b)] are similar to that of the monolayer  shown in Figs.~\ref{motif_comparison}(a,b). The STM image provides a measure of its  thickness  from  the height  profile along  the line named as {\it lmno} \textit{i.e.} the red line in Fig.~\ref{leed_stm_intrm}(a). It shows that there is considerable variation in the thickness and the maximum local height is $\approx$1.2 nm around $n$  with respect to the  region around $m$ that has lowest height [Fig.~\ref{leed_stm_intrm}(c)]. The height histogram of Fig.~\ref{leed_stm_intrm}(d)  is a nearly symmetric peak with maximum at 0.65 nm with  50\% area (red shaded) having a thickness $\ge$0.65 nm.  So,  the peak position (0.65 nm) is considered to be the average thickness of the film referenced to $m$. Since the \anc\, surface is uniformly wetted by a monolayer of Sn, the thickness  around $m$ should be at least that of a monolayer \textit{i.e.} 0.23 nm. Thus, the average thickness of the intermediate layer is at least 0.88~nm (0.65+ 0.23 nm) \textit{i.e.} \AC0.9 nm. 
	
\subsection{Sn thick films on \anc} 
\label{subsec:thickSnfilm}
	
\subsubsection{\underline {Sn deposition for 6$\leq t_d  \leq$27 min :}}

Here, we first establish that there is a continuous increase of the average thickness of the Sn film  with $t_d$. This is shown by the Sn \textit{MNN} Auger signal that increases monotonically   [Fig.~\ref{leed_stm_intrm}(e)].  Concomitantly, the substrate related Co \textit{LMM} Auger signal  decreases  and at $t_d$= 37 min, it is almost in the noise level [Fig.~\ref{leed_stm_intrm}(e)]. 

In Fig.~\ref{leed_stm_intrm}(f), the  LEED pattern for $t_d$= 6 min with a thickness of 2 nm  shows five sets of decagonal spots. The (100$\bar{1}$0) spots also develop a continuous ring (red arrow), although the sets of spots at lower reciprocal distances do not show any such ring.  At $t_d$= 12-27 min [Figs.~\ref{leed_stm_intrm}(g-i)], the spots at the lower reciprocal distances are diminished in intensity, whereas both the decagonal spots and the ring related to (100$\bar{1}$0) are visible.  The significance of the ring is discussed in the next subsection (III\,B\,2). The  video files as a function of $E_p$ corresponding to Figs.~\ref{leed_stm_intrm}(f-i)  named as ``6min", ``12min", ``17min", and  ``27min", respectively are provided in the SM~\cite{Supp}.

		\begin{figure}[!tb]
		\includegraphics[width=125mm,keepaspectratio]{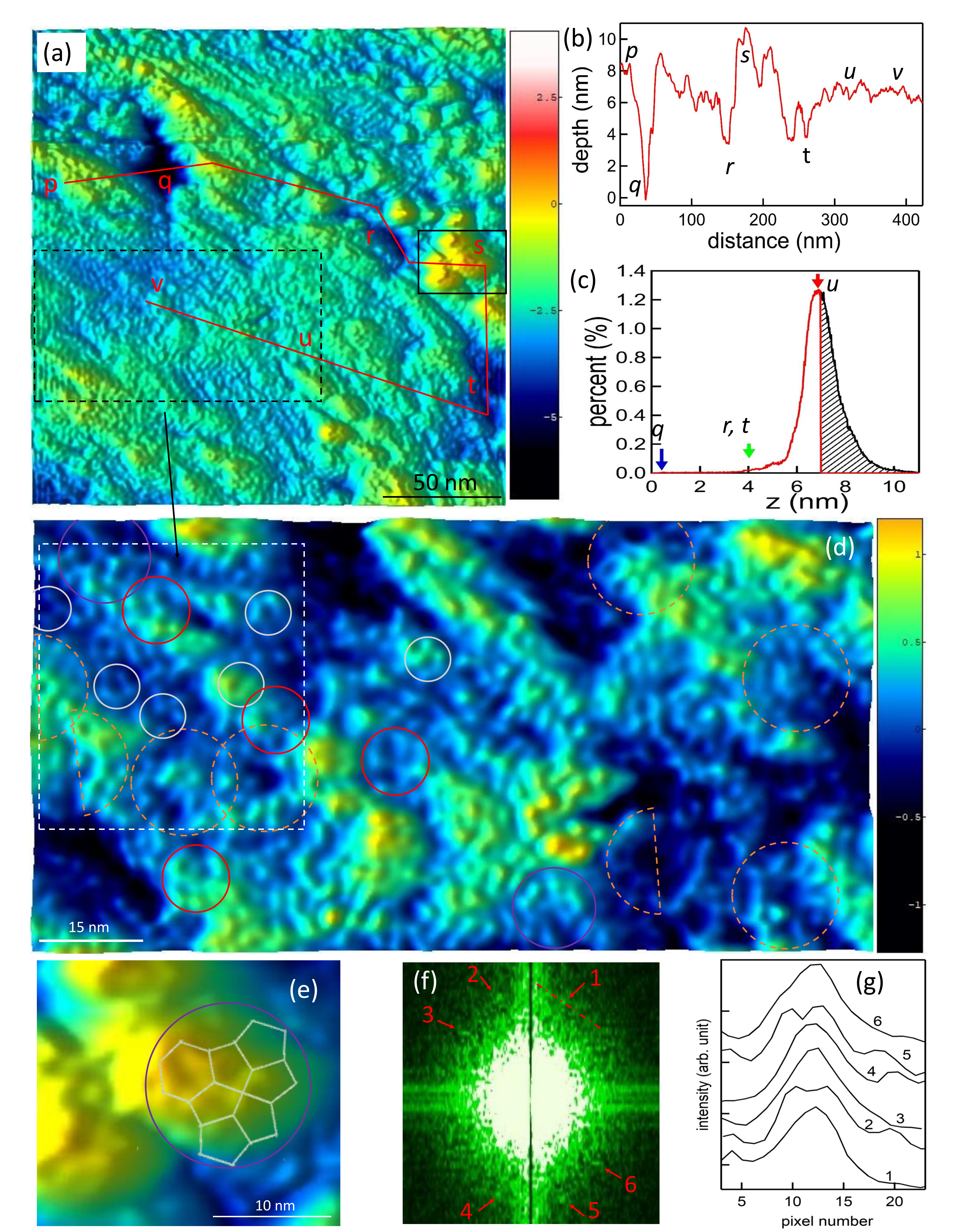}
		\caption{(a) A large area STM topography image of a  Sn thick film deposited for $t_d$= 37 min on $d$-Al-Ni-Co for LT deposition (200 nm $\times$ 200 nm, $I_T$= 0.8 nA, $U_T$= 2 V). (b) The height profile  along {\it pqrstuv} shown in red color  in  panel~\textbf{a}.	(c)  The height histogram of panel~\textbf{a}. (d) An  121 nm$\times$ 62 nm STM image from the  dashed black rectangle in panel~\textbf{a} showing motifs highlighted by colored circles as in Fig.~\ref{1ML_STM_RT}. 	These motifs are defined by and compared with  the approximant clathrate theoretical model in subsection \ref{subsec:theoryandstm}.  (e) A zoomed image   of the  largest thickness ($\approx$13 nm)  region around $s$  enclosed by a black rectangle in  panel~\textbf{a}. 	 (f) FFT of the Sn thick film with the spots indicated by red arrows and numbered  \#1- \#6. 	 (g) Intensity profiles of the  FFT spots along the tangential direction (as shown by a red dashed line for spot \#1) shown staggered along the vertical axis.}
			\label{thickSTM_LT}
		\end{figure}

\subsubsection{\underline {Thickest (10 nm) Sn film  :}}
A large area STM topography image of  a thick Sn film grown on \anc\, with $t_d$= 37 min is shown in Fig.~\ref{thickSTM_LT}(a). 	 The  height profile in Fig.~\ref{thickSTM_LT}(b) along $pqrstuv$ [red line in Fig.~\ref{thickSTM_LT}(a)] shows that the region  around $s$ (highlighted by a black rectangle)  has a height of about  10 nm with respect to the  local minimum region   (e.g., region $q$). However, the heights exhibit substantial lateral variation  \textit{e.g.} around $p$ it is 8 nm, whereas a large region around $uv$ it is 6-7 nm.  Thus, the film exhibits  a rugged topography that resembles a geographical undulated ``hilly terrain", where a valley represents the local minimum region   ($q$), but the bottom of the valley is at a larger height compared to the ``sea level" (i.e., the monolayer). 
 ~To find the thickness of the Sn film at $p$, we consider a uniform Sn deposition with time: thus for $t_d$= 37 min the estimated thickness at $q$ region is about 2.8 nm. This is estimated on the assumption that for the 0.9 nm film grown with $t_d$=  3 min, the thickness of the minimum region [\textit{m} in Fig.~\ref{leed_stm_intrm}(b)] is one monolayer (0.23 nm).  This estimate of the thickness at $q$ is consistent with the fact that the substrate Co Auger signal is almost completely suppressed [Fig.~\ref{leed_stm_intrm}(e)]. Note that the added area of regions having thickness close  to $q$ is only 1-2\% of the total area. 	 In Fig.~\ref{thickSTM_LT}(c), the height histogram shows a nearly symmetric peak centered at 7 nm ($\approx$50$\%$ of the area  shown by black shading  has a thickness $\geq$7) that is taken as the average thickness with respect  to $q$ whose thickness is 2.8 nm. Thus, by adding these two numbers, the total average thickness of the thickest Sn film studied by us  turn out to be  10$\pm$0.5 nm.   

This film  has an order of magnitude larger roughness   ($S_q$= 0.54 nm)  compared to the monolayer ($S_q$= 0.053).
 ~In spite of this, the characteristic  motifs similar to those observed for the thin films  are  observed  [Fig.~\ref{thickSTM_LT}(d)].  The largest thickness region  around $s$ portrays an elevated dome  highlighted by a violet circle in Fig.~\ref{thickSTM_LT}(e), whose average base diameter  is   12$\pm$3 nm, as shown by the histogram in Fig.~S7 of SM~\cite{Supp}. It is interesting to note that a quasiperiodic motif is  observed on the dome, which partially resembles a polygon assembly  (see \ref{subsec:theoryandstm}).

\begin{figure}[tb]
	\centering
	\includegraphics[width=165mm,keepaspectratio] {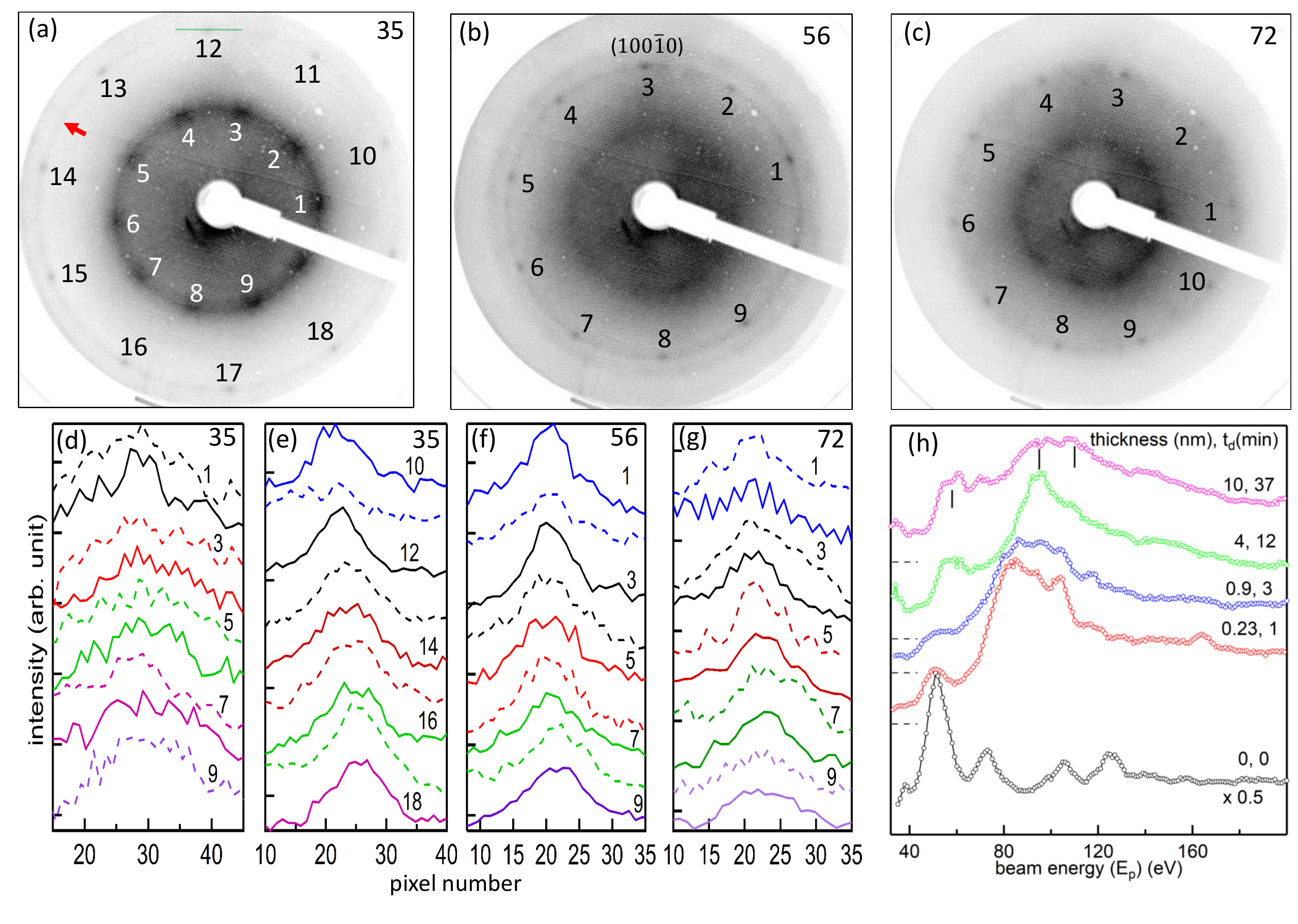} 
	\caption{LEED patterns of the 10 nm thick Sn film  recorded with $E_p$=  (a) 35 eV, (b) 56 eV, and (c) 72 eV shown in an inverted scale.  The intensity profiles 
		~along the tangential direction (\textit{e.g.} along green line in panel \textbf{a} for spot 12) through the spots (d) 1 to 9   and (e) 10-\,18 in panel \textbf{a}; (f) 1-\,9 in panel \textbf{b};  (g) 1-\,10  in panel \textbf{c}.   (h)  The I-V curves corresponding to the  (100$\bar{1}$0) spot for films of different thickness and $t_d$ shown staggered along the vertical axis, the zero for each  shown by dashed horizontal lines.  } 
	\label{thickLEED_LT}
\end{figure}

In Fig.~\ref{thickSTM_LT}(f), an averaged FFT of the STM images shows occurrence of 6 spots (\#1-- \#6). Their presence  is established in Fig.~\ref{thickSTM_LT}(g) by the intensity profiles across the spots along the tangential direction (as shown for spot 1 by a red dashed line).  It may be noted that a notional circle joining the spots is distorted to an oval shape, which could be related to the thermal drift  in the STM measurement~\cite{Singh_prr20}. The spots are weaker compared to that of the monolayer, as also in the case of the LEED  spots shown in Fig.~\ref{thickLEED_LT}. Possible reasons for this could be  
	~  the intrinsic puckered nature of the film, as shown in \ref{subsec:theoryandstm},  stochastic processes,  competing disorder, 
	~and  the roughness of the film due to growth at LT that degrades the quality of the STM images. 

 LEED investigation of the  Sn thick film  shows 10f  spots highlighted by numbering  in Figs.~\ref{thickLEED_LT}(a-c), and their intensity profiles are shown in  Figs.~\ref{thickLEED_LT}(d-g).  A peak representing each spot is clearly observed with respect to the background, and these decagonal spots    provide the  evidence of the decagonal symmetry in the thick film.  At $E_p$= 35 eV, the outer set of spots are  18$^{\circ}$ rotated with respect to the inner spots [Fig.~\ref{thickLEED_LT}(a)], and ratio of the radii of   the outer  spots (10--18) and the inner spots (1--9)  is 1.86 ($\approx$ $\tau\chi$= 1.9).  Any possibility of crystalline domains ~\cite{Sharma_jcp14}  or presence of approximant phases are ruled out because there are no extra spots or splitting of the spots at any $E_p$ (also see the  video file named ``10nm" in SM~\cite{Supp}). All the spots move towards the (0,0) spot as $E_p$ increases.  It may be noted that besides the discrete spots, as in Figs.~\ref{leed_stm_intrm}(f-i), a weak continuous ring that joins the (100$\bar{1}$0) spots is observed [red arrow in Fig.~\ref{thickLEED_LT}(a)]  possibly indicating a rotational degree of freedom between the decagonal structures. 
 	~ The latter could probably be related to its clathrate structure, since Engel $et~ al.$ proposed  possibility of random tiling quasicrystal in clathrates that are axially symmetric~\cite{qcla}. However, this would mean that the intermediate regions connecting the decagonal structures would  not be periodic. In subsection \ref{subsec:theoryandstm} (Fig.~\ref{motif_conta}), we discuss this further by comparing the STM image of the 10 nm film with  DFT based decagonal clathrate  model that is presented in the  subsection (\ref{subsec:surfmodel}).

  The I-V curve for the (100$\bar{1}$0) spot of the  Sn thick film  exhibits  peaks centered at  $E_p$=  58, 95, and 110 eV as shown by the ticks in the top curve of Fig.~\ref{thickLEED_LT}(h).   The peak positions are similar to the  4 nm film ($t_d$= 12 min), although in the latter, the 95 eV peak is somewhat more intense.  The peaks in I-V curves of the thin films  [0.23 nm i.e. one monolayer and 0.9 nm  in Fig.~\ref{thickLEED_LT}(h)], although have nearly similar  shape, are  shifted by about 8-15 eV towards lower $E_p$, which implies a possible change in the lateral length scale~\cite{Vuorinen12}. 
~  Finally, it is noteworthy that the shape  of the I-V curves of the Sn films of different thicknesses starting from the monolayer (0.23 nm) are completely different from that of  the substrate  that has the peaks  at $E_p$= 52, 73, 106, and 125 eV. This shows that the structure of the Sn films, although decagonal, is  different from the substrate.

\subsection{Model for the atomic structure, surface termination and DFT-optimized reconstruction of the Sn decagonal clathrate}
\label{subsec:surfmodel}

The low-temperature stable phase of Sn, the $\alpha$-Sn in cubic diamond $sp^3$-bonded structure (``gray tin''), transforms  to metallic and denser tetragonal $\beta$-Sn (``white tin'') above $\sim$286 K. The  clathrate structures are also  $sp^3$-bonded  by grouping atoms around point centers into empty cages. 
~In our experiments, the thick Sn film was deposited at $\sim$165K, deeply in the stability range of $sp^3$-bonded structures. At zero temperature, the cohesive energy differences of the alternative candidates for stability relative to $\alpha$-Sn obtained from DFT are:  +46 meV/atom for the $\beta$-Sn and +29 meV/atom for the face centered cubic clathrate type II  (that contains highest fraction of the smallest dodecahedral cages and has lowest energy out of all clathrates, see Ref.~\onlinecite{Singh_prr20} for a discussion about type II and III clathrates).  Given the fact that the \anc\, substrate surface exhibits excellent match with the clathrate structure (see subsection~\ref{section:nucleation}), nucleation of the $\alpha$-Sn and $\beta$-Sn 
structures must be suppressed due to the surface structure incompatibility.  Moreover,  the low deposition temperature is  another disadvantage factor for metallic $\beta$-Sn. Presumably, an {\em amorphous} Sn phase 
~might compete with the magic clathrate interfacial compatibility with \anc. But while Si or Ge amorphous phases can be prepared easily and they have apparently $sp^3$-bonding
nature, reports on amorphous Sn are very scarce, and to our knowledge literature does not report $sp^3$-bonded, low--density amorphous Sn phase that might be a hypothetical competitor at low temperatures. 

In addition to these general considerations, the clearly recognizable \textit{decagonal} symmetry of the LEED  pattern from the thick Sn layer rules out any of the alternatives considered above.  The role of the substrate is to  prevent possible formation of crystalline Sn structures.  In the following, we develop detailed theoretical -- although indirect -- support for {\em decagonal clathrate} as the only plausible candidate model structure for the thick Sn film.  In the following subsection (\ref{subsubsec:structure}), we argue how  
the decagonal clathrate follows from the dual relationship (where connecting the centers of the faces of one structure gives the other e.g., an icosahedron gives a dodecahedron) between the Frank--Kasper and the clathrate structures~\cite{Frank_58,Frank_59,Okeeffe_pml98}. We then review the decoration prescription associating the precise bulk clathrate atomic structure of Sn with two-dimensional geometry of the R-T tilings. The next
subsection~\ref{subsubsec:terminations} discusses the decagonal clathrate surface terminations, and finally subsection~\ref{subsubsec:reco} reports on the DFT--guided energy optimization of such terminations by adding adatoms  at specific locations to neutralize the energy cost of the unsaturated bonds at the surface.

\subsubsection{\underline {Tiling description of the decagonal clathrate:}}
\label{subsubsec:structure}
\begin{figure}
		\includegraphics[width=165mm,keepaspectratio]{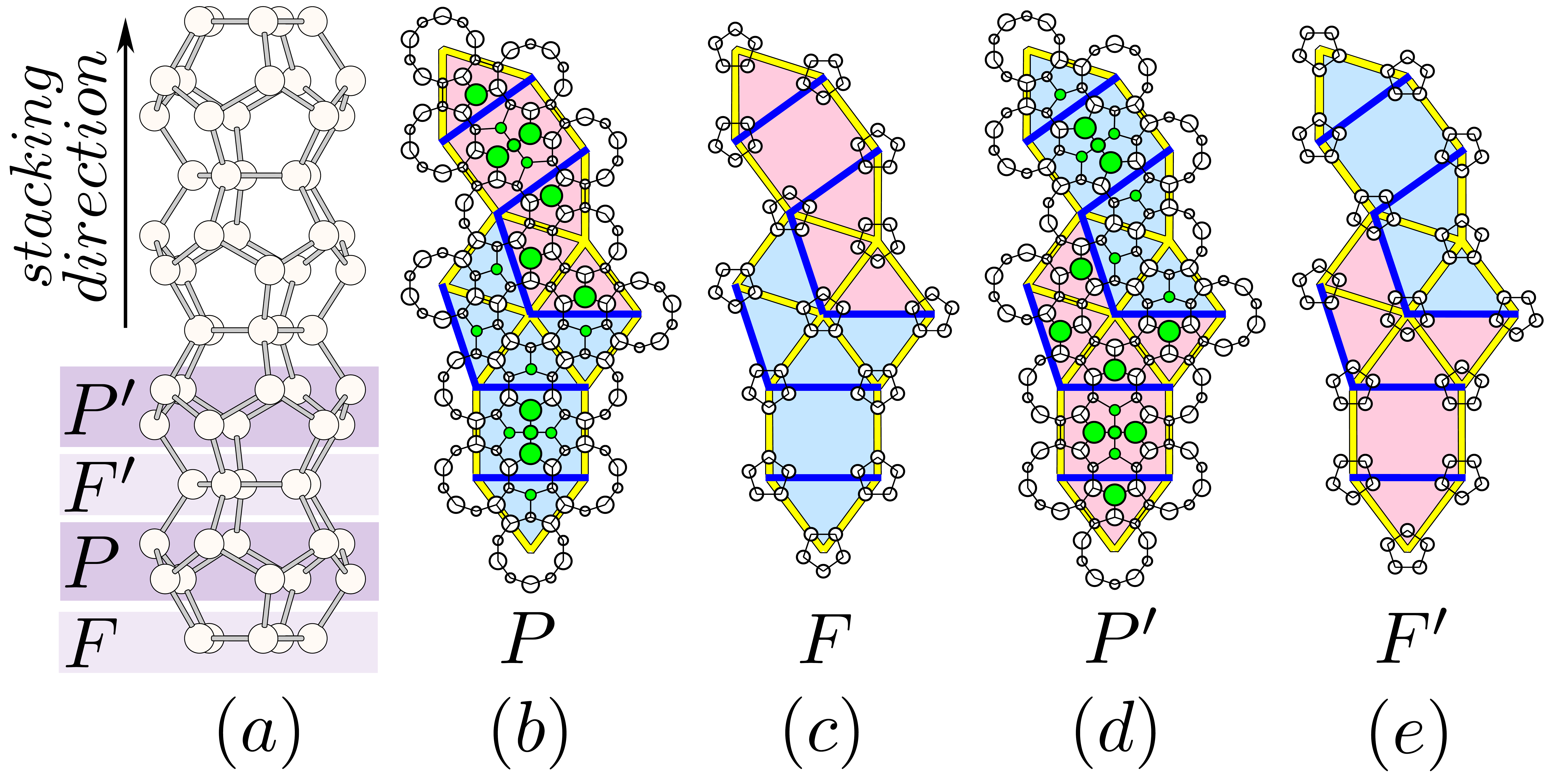}  
			\caption{ (a) Side view of the column of stacked dodecahedra, decorating	the	vertices of the R-T tiling, the stacking  direction $z$ perpendicular to the substrate surface is indicated by a vertical arrow, its length corresponds to the stacking period $c$ (= 1.251~nm).  Top view of the layer-by-layer decomposition of one repeat ($PFP'F'$) of the bulk structure for a small patch of the R-T tiling, where the layers are denoted by (b) $P$, (c) $F$, (d) $P'$, and (e) $F'$. Open circles are Sn atoms that belong to the columns	of the dodecahedra. Green circles are ``interstitial'' Sn atoms outside the	columns of stacked dodecahedra. The sizes of the circles representing the atoms scale with their vertical height.}
		\label{fig:laybylay}
	\end{figure}
	
	Frank and Kasper~\cite{Frank_58,Frank_59} pointed out that the network of clathrate cage centers forms tetrahedrally close-packed (TCP) geometry, and suggested that new clathrate structures can be derived from the known examples of the metallic structural family. Following this route, a decagonal TCP quasicrystal structure~\cite{Mihalkovic_14} can be dual-transformed into {\em decagonal clathrate}~\cite{Singh_prr20}. The common geometrical framework of the two related structural families is decagonal tiling of rectangles and isosceles triangles (R-T tiling hereafter). 

The decagonal R-T tiling -- common geometry for metallic TCP	structures or clathrates -- consists of  isosceles triangles (T)	with two shorter sides with length $a$ and a longer side	$b$= $a\sqrt{\tau+2}/\tau$  and rectangles (R) with aspect ratio of $\sqrt{\tau+2}/\tau$. In a random R-T tiling, R or T tiles pack without restrictions by sharing $a$ or $b$ edges.  DFT--optimized tile edge lengths for the clathrate structure are $a$= 1.082 nm 
~ and	$b\approx$1.268~ nm. 
	~ Wider angle of the T tiles is $2\pi/5$ so that five T's pack around a common vertex implementing a local 5--fold symmetry. R-T tiling is {\it bipartite}, since tiling vertices can be uniquely partitioned into ``even'' and ``odd'' via a simple rule: pairs of vertices connected by $a$--type linkage must have opposite parity. 
~	Decagonal R-T tiling has a close relationship with tilings of Ammann rhombuses: T tile is exactly half of the ``fat'' rhombus, and the R tile connected via $b$-type edge with T tile corresponds to a pair of 36$^\circ$ skinny rhombs, plus a T tile. 
		
	Given an R-T tiling, the backbone of the decagonal clathrate structure ($\sim$ 85\% of the atoms) is given just by columns of dodecahedral cages centered around tiling vertices, as illustrated in 	Fig.~\ref{fig:laybylay}(a). 
  ~The dodecahedra forming the column are stacked on the top of each other, defining vertical stacking period of
	$c= 2d_{dod}\approx$1.251~nm, where $d_{dod}$ is the shorter face-to-face diameter of the dodecahedron. The columns possess {\it local} 10--fold screw	symmetry axes $10_2$, and the nearby columns related by translation over $a$--type edges are mutually rotated by 2$\pi/10$, implementing the bipartitness property.

	The dodecahedron columns can be also viewed as stackings of  ``puckered'' 10--rings and ``flat'' pentagons.
	Figs.~\ref{fig:laybylay}(b-e) illustrate how this column decomposition invokes {\em layering} of the structure: the 10--rings	give rise to strongly puckered $P$--type layers, while pentagons are always located at singular height in ``flat''  $F$--type layers.  Full stacking period then reads \textbf{$FPF'P'$}, 	where the primed layer  (10--rings or pentagons) are related to the non-primed motifs by action of the local $10_2$ screw axis associated with every vertex column: 2$\pi/10$--rotation combined with half--period translation. 	The four stacking--period layers are centered at fractional heights $z$= 0 ($F$), 1/4 ($P$), 1/2 ($F'$), and 3/4 ($P'$), respectively. While $F$--type layers occur at single discrete heights, the $P$--type layers span significant width of 3.2\,\AA, measured from the top most to	the bottom most $P$--layer atom.  Separation from a topmost $P$--layer 	atom to the $F$--layer plane is 1.5\,\AA.  
	
	The ``decoration rule'' is completed by placing ``interstitial'' atoms	outside the dodecahedral columns; these are shown as green filled	circles in the Fig.~\ref{fig:laybylay}(b-e). Upon placing the interstitial atoms, every Sn atom in the structure satisfies	$sp^3$-bonding constraints: the coordination is strictly four, and the	bond-angles vary within 105-120$^\circ$ (compared to 	$\approx$109.5$^\circ$ bond angles in the $\alpha$-Sn diamond structure).
	
	The symmetry relationships between local atomic motifs are conveniently	represented by tile coloring: both R and T tiles are either pink, or light-blue, depending on the orientation of pentagons in the $F$ and $F'$	layers. $P$/$P'$ layer projected atomic positions are the same for 	either tile coloring, but their {\em vertical displacements} from the $z= 1/4$ or $z= 3/4$	 fractional heights [reflected by circle sizes in	Fig.~\ref{fig:laybylay}(b-e)] {\em are opposite} for the pair of colored tiles.

	\begin{table}
	\begin{center}
		\begin{tabular}{ccccc}
			& \multicolumn{2}{c}{triangle \textbf{(T)}} & \multicolumn{2}{c}{rectangle \textbf{(R)}}\\
			& $F$ & $P$ & $F$ & $P$ \\
			\hline
			N$_{at}$ & 5/2 & 6 & 5 & 15\\
			$\rho_\alpha$[ML$_\alpha$]& 0.4507 & 1.0816 & 0.3646 & 1.0938\\
		\end{tabular}
	\end{center} 
	\caption{\label{tab:tiles}
		The atomic content of R and T tiles  averaged over primed and non-primed layers. 
		  ~$N_{at}$ gives the number of Sn atoms per respective tile, 		and $\rho_\alpha$ is the coverage expressed in units of the (111)-type $\alpha$-Sn monolayer ML$_\alpha$.} 
\end{table}
	
	Table~\ref{tab:tiles} summarizes the atomic content of the R and T tiles. In case of decagonal tiling, number frequency ratio of the T	to R tiles is 4$\tau$, and the respective tile areas are $S_{\rm T}=a^2\sqrt{\tau+2}/4$ and $S_{\rm_R}=a^2\sqrt{\tau+2}/\tau$; hence coverage contribution $\rho_\alpha$ in the {\em decagonal} structure will be 0.427 
	 ~ML$_\alpha$ for $F$--type, and 1.085 ML$_\alpha$ for $P$--type layers. 

	\begin{figure}
	\includegraphics[width=100mm,keepaspectratio]{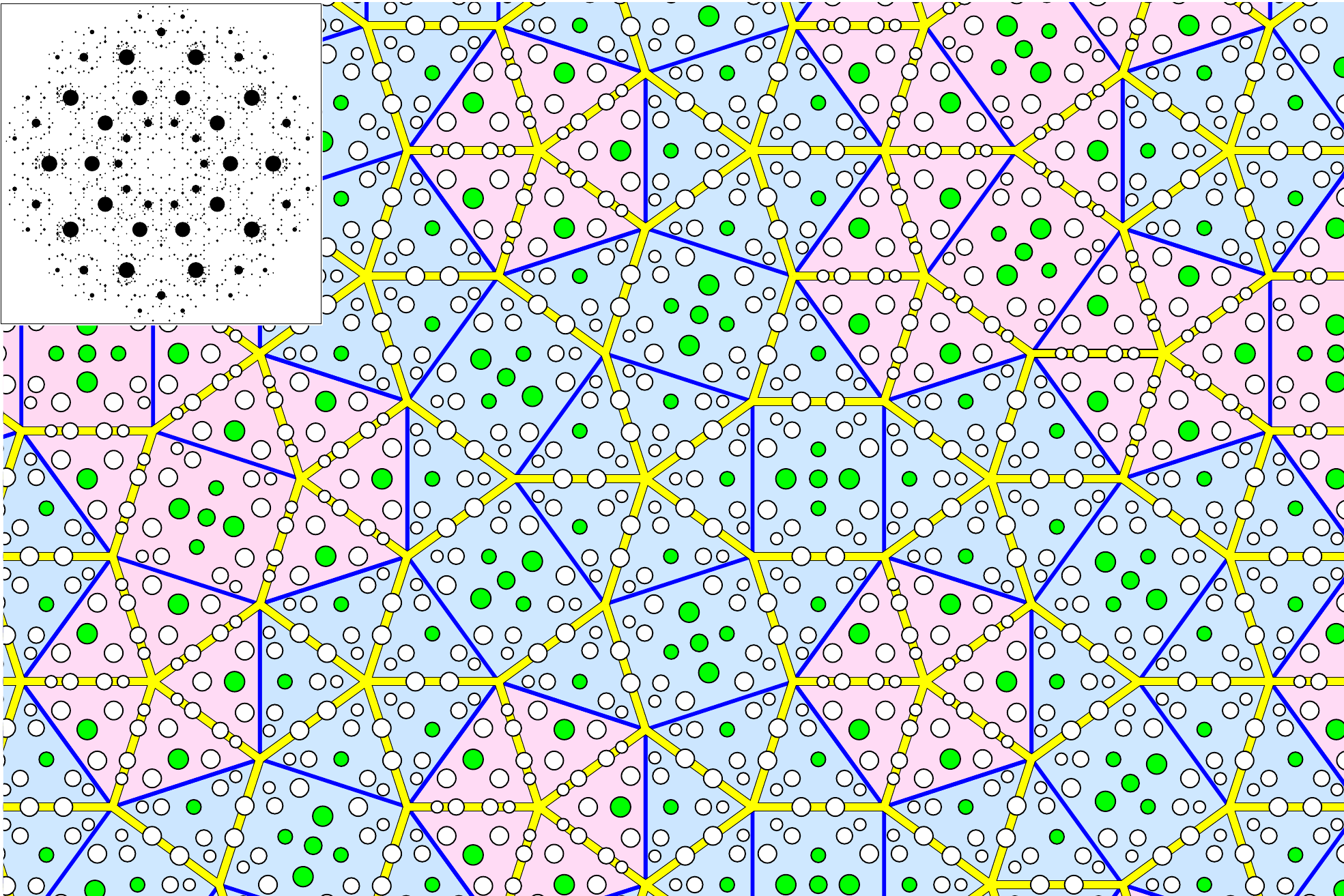} 
	\caption{Fragment of a large decagonal approximant with 2692 atoms per cell	showing a $PF$--bilayer slice, surface perpendicular to the pseudo-10-fold axis.  Tile coloring as in Fig.~\ref{fig:laybylay};  also note that superposition of Figs.\ref{fig:laybylay}(b,c) gives the bilayer slice shown here for a much bigger continuous tiling fragment. The inset shows the diffraction pattern of the approximant structure parallel to the stacking $c$ direction, revealing nearly perfect 10-fold symmetry.}  
	\label{fig:structure}
\end{figure}

	In their pure forms, the decorated R and T tilings exactly correspond	to the well known canonical clathrates  composed from four canonical cages, entirely filling the space.
	 ~In particular, the	pseudo--10-fold axis is parallel to the cubic (110) direction in	clathrate Type II structure revealing isosceles triangles geometry, and to	(100) direction in type III hexagonal clathrate showing pure-R	tiling.  Fig.~\ref{fig:structure} shows a large decagonal clathrate	``approximant'' with 10.25\,nm$\times$\,8.72\,nm$\times$\,1.25\,nm~sides 
	~of the periodic cell, in which R and T tiles pack together to form	eventually large pentagonal motifs. Inset in the upper left corner of	the panel shows (pseudo)-10--fold zeroth-layer of the diffraction pattern -- deviation from perfect decagonal symmetry is hardly visible.

	\subsubsection{\underline {Decagonal clathrate -- surface terminations:}}
	\label{subsubsec:terminations}
	We require that a proper Sn clathrate termination -- in analogy with $\alpha$-Sn diamond surfaces -- leads to at most one unsaturated bond for any atom	exposed on the surface, \textit{i.e.} the coordination number is $\geq 3$. Optimization strategy based on this assumption	proved successful in predicting clathrate structures of reconstructed,	free-standing ultra-thin slabs as a ground--state between 2.5-7 ML$_\alpha$
	thickness for Si and Ge, and for 5--8 ML$_\alpha$ thick Sn slabs~\cite{Pospisilova_prep}. Here, we define ``thickness'' or {\em coverage} in units of (111)-type monolayer of the diamond structure ($\alpha$-Sn in case of tin). Note that  our experimental film thickness is derived from 
	~the height coordinate, hence it is not directly connected with ML$_\alpha$ units used in this section.
       
    \begin{figure}
	\centering
	\includegraphics[width=.95\textwidth]{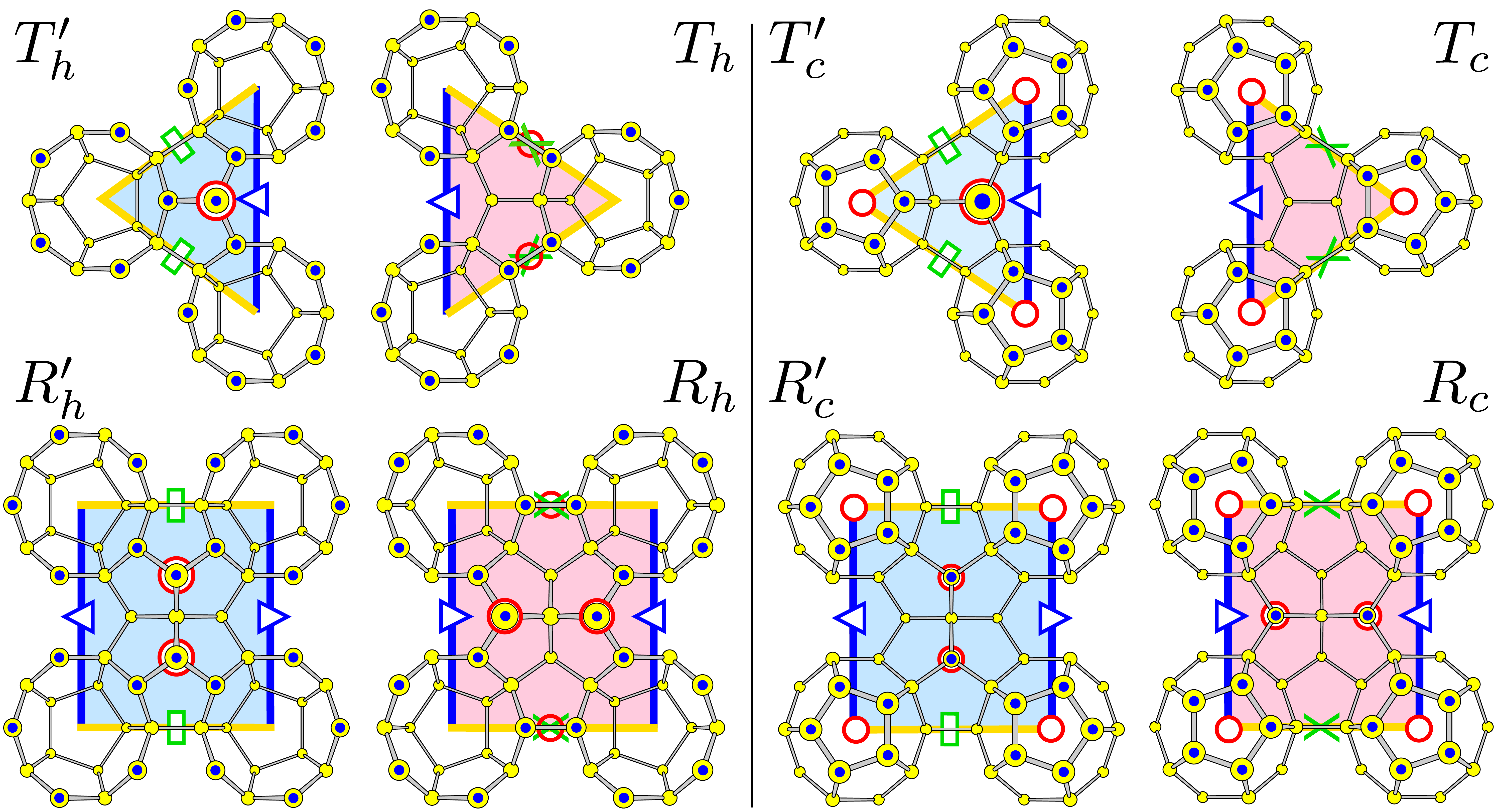}
	
	\caption{Tile-decoration rule for T and R tiles shown in the top view. Edge markings (green crosses and rectangles, blue triangles)	implement ``packing rule'' for edge-sharing in a tiling.   The tiles on the left side of the vertical line with terminations just above $P$ type layers with 10-rings are	subscripted with letter ``{\em h}'' (hole), while those on the right with	$PF$ type termination with ``penta-caps'' on the top of 10-rings are labeled with subscript ``{\em c}''. All Sn atoms are represented by yellow circles, atoms marked by	blue circular spots in the center are 3--coordinated (before	reconstruction). Red	circles mark energy minimizing adatom positions. } \label{fig:tiles}
\end{figure}
	
	It turns out that straight planar cuts normal to the stacking direction are valid surface terminations, satisfying the
	coordination-number requirement, for cuts at heights in	between layers defined in Fig.~\ref{fig:laybylay}. Atoms exposed to the 	surface are always from {\it two} topmost layers, $P+F$ or $F+P'$ or in short  $PF$ or $FP'$ layers. 
        ~	Termination by $F'P$ or $FP'$ bilayers with $P$/$P'$ on the top	exhibits 10--rings centered on the stacked dodecahedra column axis -- at	the tiling vertices. 
	~Termination by $PF$ or $P'F'$ bilayer (with topmost $F$ layer) places pentagonal caps on the top of the 10--rings, thus closing the dodecahedral cages at	the surface. 
    
	Fig.~\ref{fig:tiles} depicts the two possible	terminating surfaces for each kind of the tile on the surface: T or R tiles with topmost layer $P$ are designated $T_h$/$T'_h$ and $R_h$/$R'_h$, respectively for the two flavors of each tile. 
	~Atoms represented by  yellow circles with a blue circular spot at the center are 3--coordinated atoms that possess one unsaturated bond, each. The edge markings (green 	crosses/ rectangles for yellow $a$-type linkages, blue triangles for	$b$-type linkages) implement tile packing rules for valid clathrate 	decorations.
	
	{\it Comparison with diamond structure (111)-type termination and clathrate growth:}  The unreconstructed (111)-type diamond surface of Sn shows puckered even--odd honeycomb lattice with even/odd sites	located at two nearby planes, that are $\sim$1~\AA\, apart from each	other. The topmost sublayer exposes 3--coordinated atoms while 	bottom sublayer atoms are 4--coordinated. Hence, 50\% of the atoms on	the surface have unsaturated bonds.
	
	In case of clathrate Sn, $P$ or $P'$ terminating layers provide 1.09 ML$_\alpha$ 
~		coverage (where ML$_\alpha$ refers to diamond-111 surface layer), out of	which 0.53 ML$_\alpha$ atoms are 3--coordinated, the fraction	0.53/1.09$\approx$0.486 is very similar to the diamond structure.  However, capping the 10--rings by 5-caps in another 	growth stage adds another 0.43 ML$_\alpha$ coverage, while {\em keeping the	number of 3--coordinated atoms constant}. Thus the $PF$ (or $P'F'$) bilayers terminated by 5-caps of the $F$--type layers add 1.52 (1.09+ 0.43) ML$_\alpha$ of the coverage, out of which less than 35\% of atoms have unsaturated bonds -- appreciably less than in the case of diamond. Width of one such bilayer is $\gtrsim$6\,\AA, about the same as two (111)-layers of  the diamond structure.
	
	Since the 10-ring centers are laterally separated by $a$ or $b$ linkages 	($\sim$1-1.2 nm) shown as yellow or blue tile edges in	Fig.~\ref{fig:laybylay}, capping the 10--rings by 5-caps proceeds independently from one column of dodecahedra to another, which presumably precludes ``layer-by-layer'' growth mode. This is  not surprising because of the 3D character of the clathrate fundamental units - cages. 	On the other hand, cages can be viewed as a local, relatively stable unit in which Sn atoms are not less than 3--coordinated even when the	cage is entirely isolated. The unsaturated bonds on the cage surface then act effectively as attractive potential, coalescing the cages	together while saturating the bonds, and minimizing the surface area during	growth. We expect that the growth mode of the Sn clathrate will be either pure island or	mixed island--layer type, depending on the energy scales related to	processes of (1) completion of a surface cage and  (2) coalescing cages	together.
	
	To summarize, in our idealized model the clathrate growth proceeds in two stages: building mutually interconnected network of the $P$ layer	atoms exhibiting nearly equal number of unsaturated bonds per surface atom as the diamond structure (111)-surface, while in the second stage the 10--rings are covered by 5-caps in a stochastic process proceeding	randomly/independently at any location on the surface.
	

	\subsubsection{\underline {Surface reconstruction}}
	\label{subsubsec:reco}
	
	    Terminations of $sp^3$-bonded surfaces necessarily produce unsaturated dangling bonds with significant energetic cost; the latter can be reduced 
	    ~by appropriately positioned ``adatoms'' that do not occupy generic lattice positions. In this subsection, we explore adatomic modifications of the clathrate surface structure by minimizing DFT surface energies.

	\begin{figure}
		\centering
		\includegraphics[width=120mm,keepaspectratio] {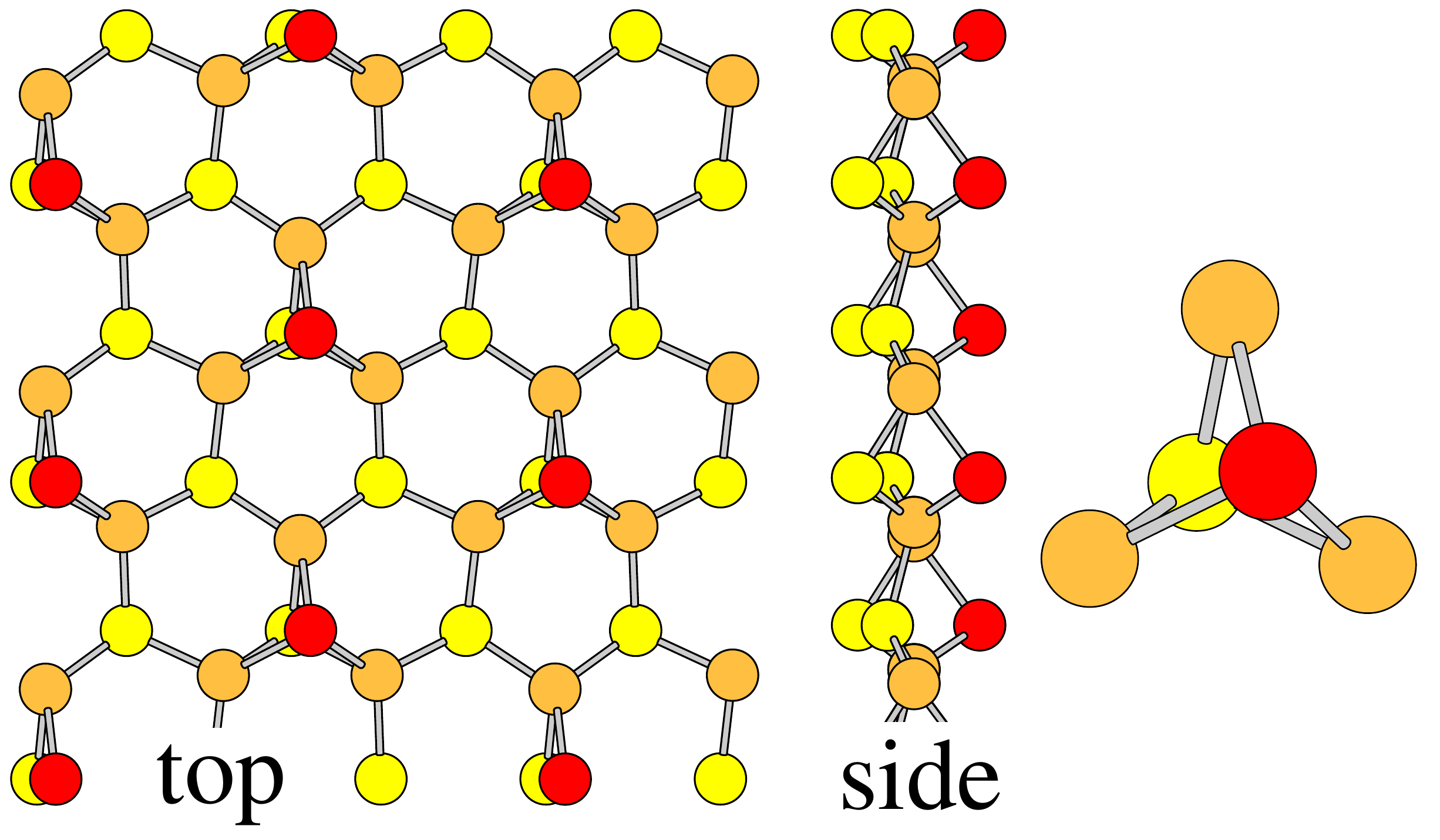}
		\caption{ Top and side views of the $\sqrt{3}\times\sqrt{3}$-reconstruction of the diamond $\alpha$-Sn(111) surface: adatoms are shown in red, 3--coordinated Sn atoms  (become 4--coordinated {\em after} adatom placement) are orange, and fully 4--coordinated Sn atoms are yellow. Yellow atoms represent  the topmost bulk layer. {\it Right:} the dumbbell motif.}
		\label{dia111}
	\end{figure}

As an example, let us first consider (111)-type surface termination of 	the related $sp^3$-bonded $\alpha$-Sn diamond structure, illustrated 	in Fig.~\ref{dia111} on the example of so called $\sqrt{3}\times\sqrt{3}$-reconstruction. The unreconstructed surface (i.e., without the red colored adatoms)	is built up from 3--coordinated orange-colored Sn atoms in the upper half-layer of the honeycomb lattice, while the yellow 4--coordinated	Sn atoms appear $\sim 1$ \AA~ below. The yellow/orange coloring divides the underlying honeycomb lattice into even/odd parts. Its	surface energy $\Gamma$ decreases from 41 to 33 meV/\AA$^2$ after positioning the Sn ``adatoms'' above some of the 4--coordinated atoms	-- they are shown in red color in Fig.~\ref{dia111}, and the ``top'' 	view is slightly off the surface normal to make the yellow atoms	visible.  The placement of adatoms leads to formation of 5--atom {\em		dumbbell} motifs enlarged in the rightmost panel of the figure. The remarkable aspect of this scenario is that the adatom (red in the	figure) forms {\em lateral}, 2.9~\AA~long bonds with three nearby 3--coordinated orange Sn atoms, while being positioned 3.4~\AA~above	the 4--coordinated yellow atom, leading to three $\sim$60$^\circ$ bond--angles between the orange--red and orange--yellow Sn atom bonds. 	Apparently, the advantage of saturating the lateral bonds exceeds the	disadvantage from introducing the $\sim$60$^\circ$ bond--angles.	
	
	Clathrate surfaces do not offer any exactly analogous configurations of the 3--coordinated atoms on the surface. Since majority of the	facets bounding the clathrate cages appearing on the surface are \textit{pentagons}, the lattice of the surface clathrate atoms is {\em not} bipartite 
~	and energy optimization requires systematic search over the	catalog of adatomic candidate sites.  Our DFT--optimized catalog of added reconstruction sites is shown in Fig.~\ref{fig:tiles}, with site's projected positions indicated by red open circles. These sites always reside about 3.4\,\AA~{\em below} a 3--coordinated atom protruding outward from the surface layer, and they fall into four	classes:\\
	\indent	1.~ below the center of protruding pentagonal  cap, for all $c$-subscripted T and R tiles. The sites appear {\em	inside} the closed dodecahedron cage at the surface;\\
	\indent 	2.~		near the center of the {\em primed} T tiles.  These sites occur inside largest, 28--atom cage, and in the case of T$^{'}_c$ triangle that has three nearby 3--coordinated	neighbors, they are most reminiscent of the diamond-structure surface dumbbell configuration;\\
	\indent 3.~ pairs of sites inside the R tiles;\\
	\indent 4.~``mid--edge'' sites in R$_h$ and T$_h$\\

	 The surface energy calculation formally attributes increased energy relative to the bulk to the area of the open surface:
	\begin{equation}
		\label{eq:esurf}
		\Gamma = ( E_{cell} - N_{cell} E_{bulk} ) / A
	\end{equation}
	where $E_{cell}$ is energy per periodic cell, $N_{cell}$ is the number of	atoms within the cell, $E_{bulk}$ is the energy per atom of the bulk structure, and $A$ is the surface area of the cell.

	However, surface energies of the tile terminations in Fig.~\ref{fig:tiles} cannot be computed directly, since even pure tilings of R or T tiles necessarily combine at least two types of the tiles -- while the sample gives us single energy. Also, in order to weaken bias from having just one particular way of joining tiles together, more alternative tile-tile arrangements are necessary. 	In Table~\ref{tab:gamma}, we report surface energies $\gamma_{j}^{tile}$ (defined by  Eq.~\ref{eq:esurf}) of eight pure decagonal tiles (shown in Fig.~\ref{fig:tiles}), resulting from least-squares fit according to Eq.~\ref{eq:gamma} for a collection of 15 slab samples labeled by $i$:
			\newcolumntype{C}[1]{>{\centering\let\newline\\\arraybackslash\hspace{0pt}}m{#1}}
	
	\begin{table}
		\centering
		\renewcommand{\arraystretch}{1.3}
		\begin{tabular}{!{\vrule width 1.5pt} C{1.0cm} | C{1.0cm} | C{1.0cm} | C{1.0cm} | C{1.0cm} | C{1.0cm} | C{1.0cm} | C{1.0cm} !{\vrule width 1.7pt}}
			\noalign{\hrule height 1.6pt}
			$T_h$  &  $R_h$ &  $T^{\prime}_h$ &  $R^{\prime}_h$ &  $T_c$ &  $R_c$  &  $T^{\prime}_c$  &  $R^{\prime}_c$  \\
			\hline
			23.1   &  23.3  &   22.4          &  22.5           &  23.3  &  22.2   &  26.4            &  25.9            \\
			\hline
			\noalign{\hrule height 1.2pt}
		\end{tabular}
		\vspace{0.25cm}
		\caption{\label{tab:gamma}
			Eight surface energies ($\gamma^{tile}_j$ in meV/\AA$^2$) for each kind of reconstructed tile in Fig.~\ref{fig:tiles}, fitted to Eq.~\ref{eq:gamma}.
		}
	\end{table}
		\begin{equation}
		\label{eq:gamma}
		\Gamma_i = \sum_{j=1}^{8}f_{ij}\gamma^{tile}_j, \qquad f_{ij} = \frac{N_{ij}s_j}{A_{i}} , \qquad i \in \{1,\,\dots,\,15\},\,\,\,\,j\in \{1,\,\dots,\,8\} 
	\end{equation}
	where $\Gamma_{i}$ is the surface energy of $i$-th tile combination, $f_{ij}$ is the fraction of $i$-th sample surface $A_{i}$ occupied by $j$-th pure tile. More precisely $N_{ij}$ is the count of tile $j$ in sample $i$, its area $s_j$'s are either $S_R$ for R tile or $S_T$ for T-tile (see 	subsection~\ref{subsubsec:structure} for definition); and	$\gamma^{tile}_j$ are the eight fitted parameters -- tile surface	energies. The samples are based on three independent tilings, pure R, pure T, and R$_2$T$_4$.
~	 The terminations based on the latter tilings	are discussed in detail below.  Unit cells of the pure tilings must
	include pairs of tiles (see edge-matching rules in Fig.~\ref{fig:tiles}): R$'_{\alpha}$R$_{\alpha}$ or
	T$'_{\alpha}$T$_{\alpha}$, where $\alpha$ stands for ``$c$'' or ``$h$''	vertex-column terminations. Bulk energies entering Eq.~\ref{eq:esurf} are $+$28.8, $+$45.6 and $+$42.9 meV/atom for T, R and R$_2$T$_4$	tilings respectively, relative to bulk energy of $\alpha$-Sn (for comparison, in our DFT setup tetragonal $\beta$-Sn is $+$45.7 meV/atom).  The samples -- slabs and bulk -- were relaxed until forces	did not exceed 0.05 eV/\AA. The ``measured''  surface energies $\Gamma$	ranged between 22-25.5 meV/\AA$^2$. The fitted surface energies $\gamma^{tile}$ are nearly 
	~similar around 23 meV/\AA$^2$, with the 	exception of T$'_c$ and R$'_c$ with
	$\gamma^{tile}_7\sim\gamma^{tile}_8\sim$26 meV/\AA$^2$.  The standard deviation is 0.4~meV/\AA$^2$, and maximal data-fit discrepancy is 0.65 in the same units. The range of	the surface energies is comparable to that of  	hexagonal $\gamma$-Sn	surfaces reported in Table II of Ref.~\cite{gammaSn}:  $\approx$21--28 meV/\AA$^2$,   and appreciably less than	the $\alpha$-Sn (111)-type surface in ($\sqrt{3}\times\sqrt{3}$)--type 	adatomic reconstruction  (33 meV/\AA$^2$).

\begin{figure}
		\centering
		\includegraphics[width=70mm,keepaspectratio]{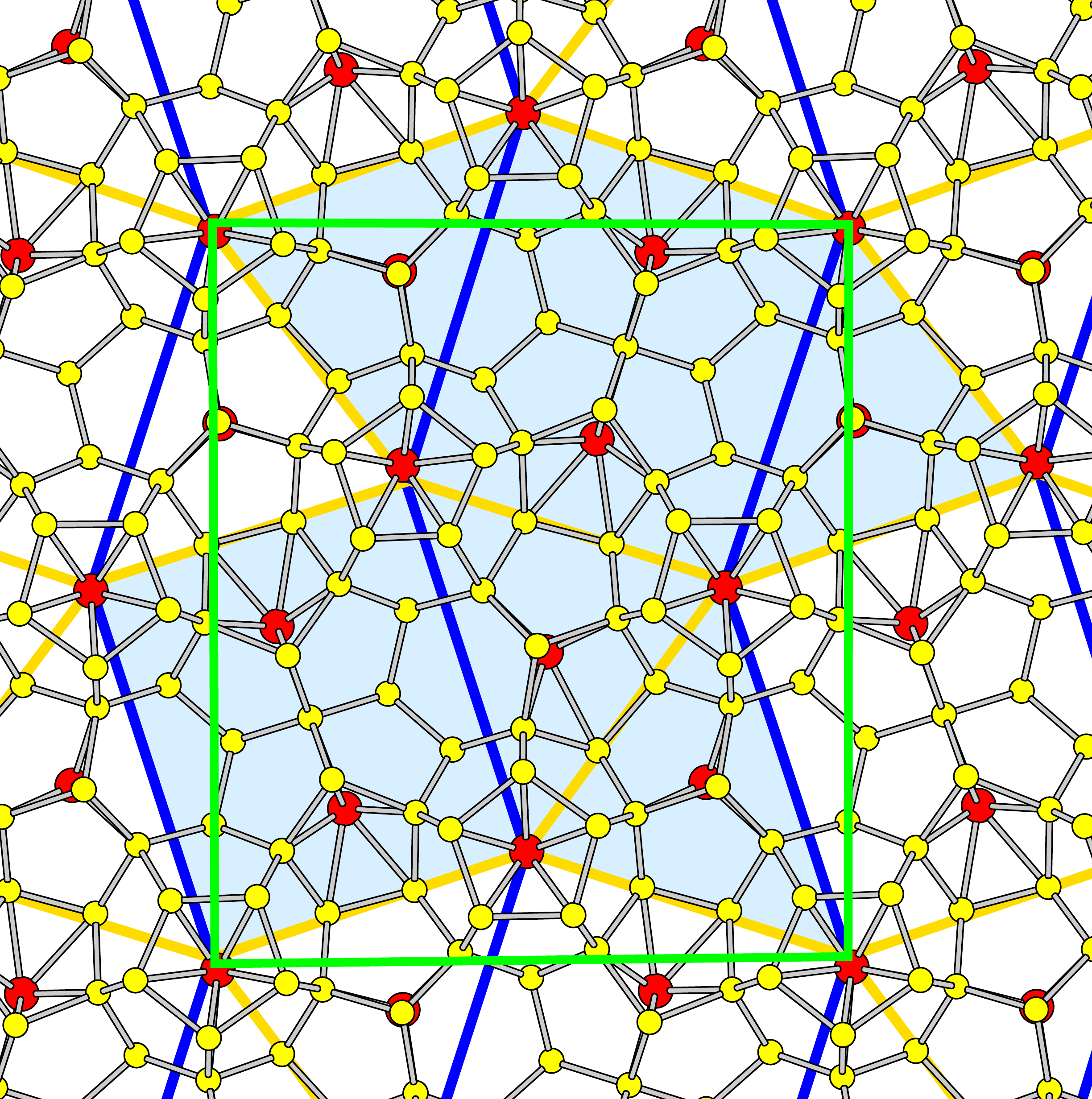}
		\hspace{0.5cm}
		\includegraphics[width=70mm,keepaspectratio]{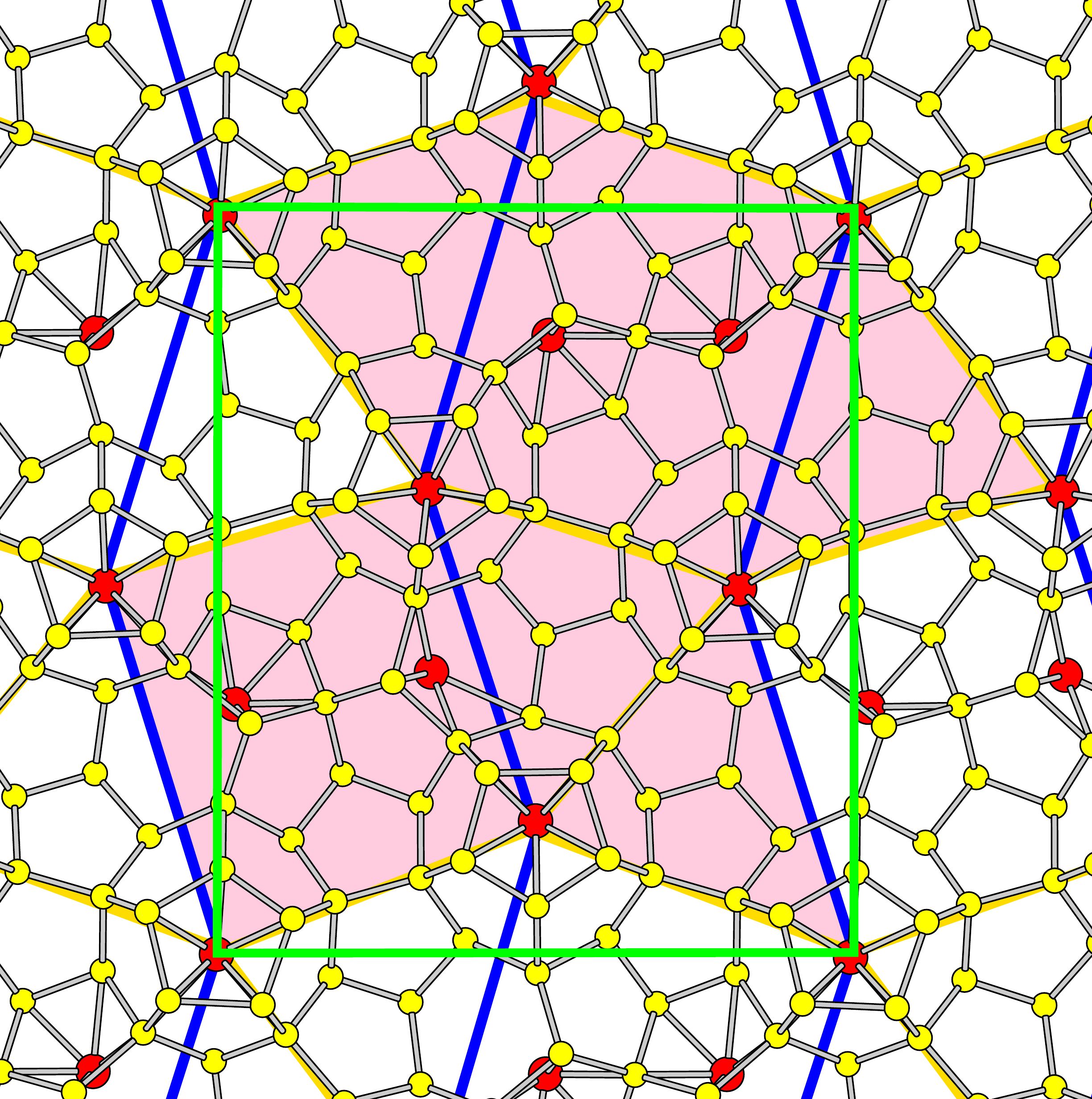}
		\caption{4T$'_c$+2R$'_c$ ({\em left}) and  
			4T$_c$+R$_c$ ({\em right}) terminations of the R$_2$T$_4$ clathrate structure. Adatoms are shown
			by red circles, a green rectangle outlines the periodic boundary in $ab$ plane.   	}
		\label{penta-caps}
	\end{figure}

\begin{figure}
		\centering
		\includegraphics[width=70mm,keepaspectratio]{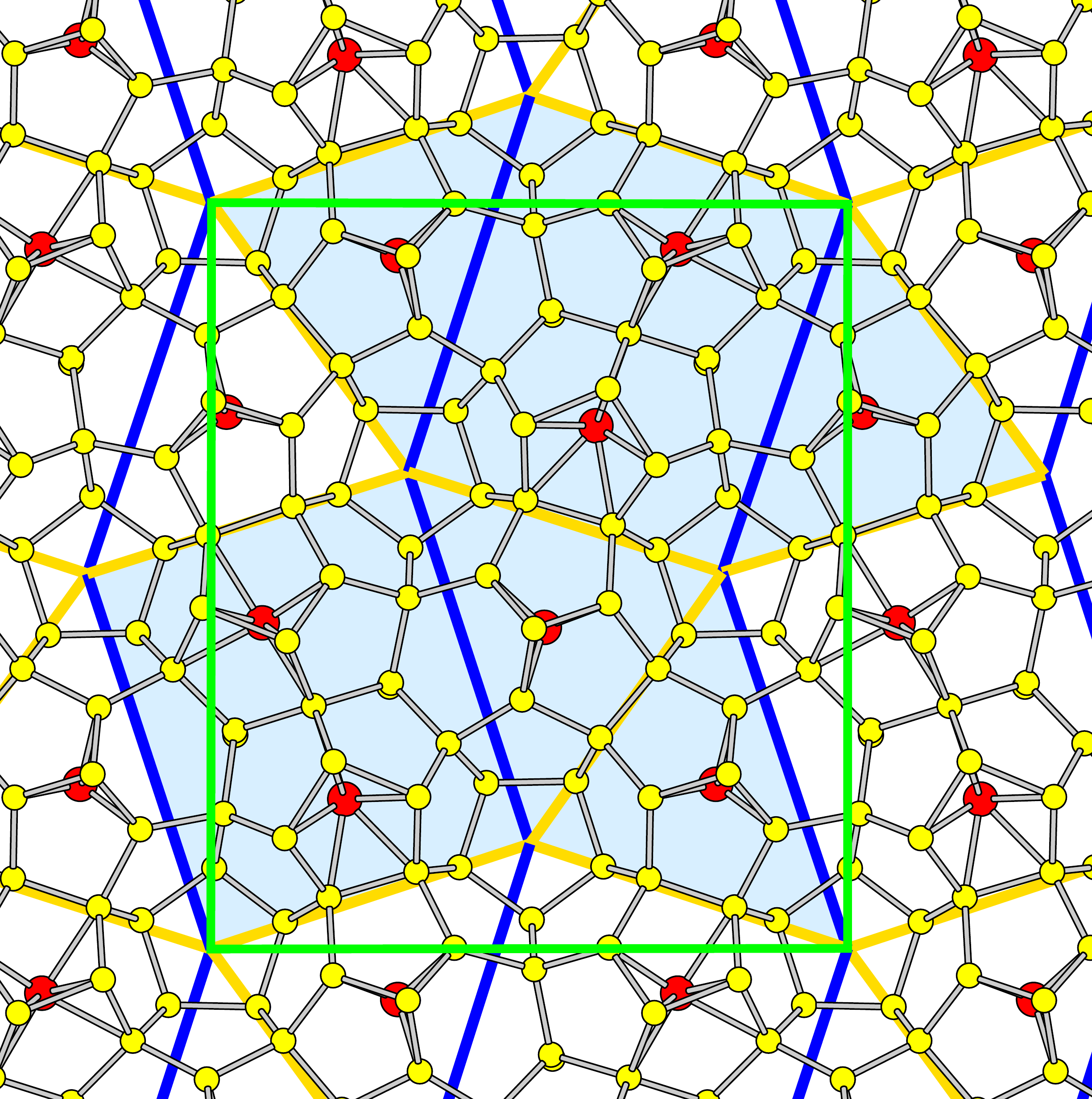}
		\hspace{0.5cm}
		\includegraphics[width=70mm,keepaspectratio]{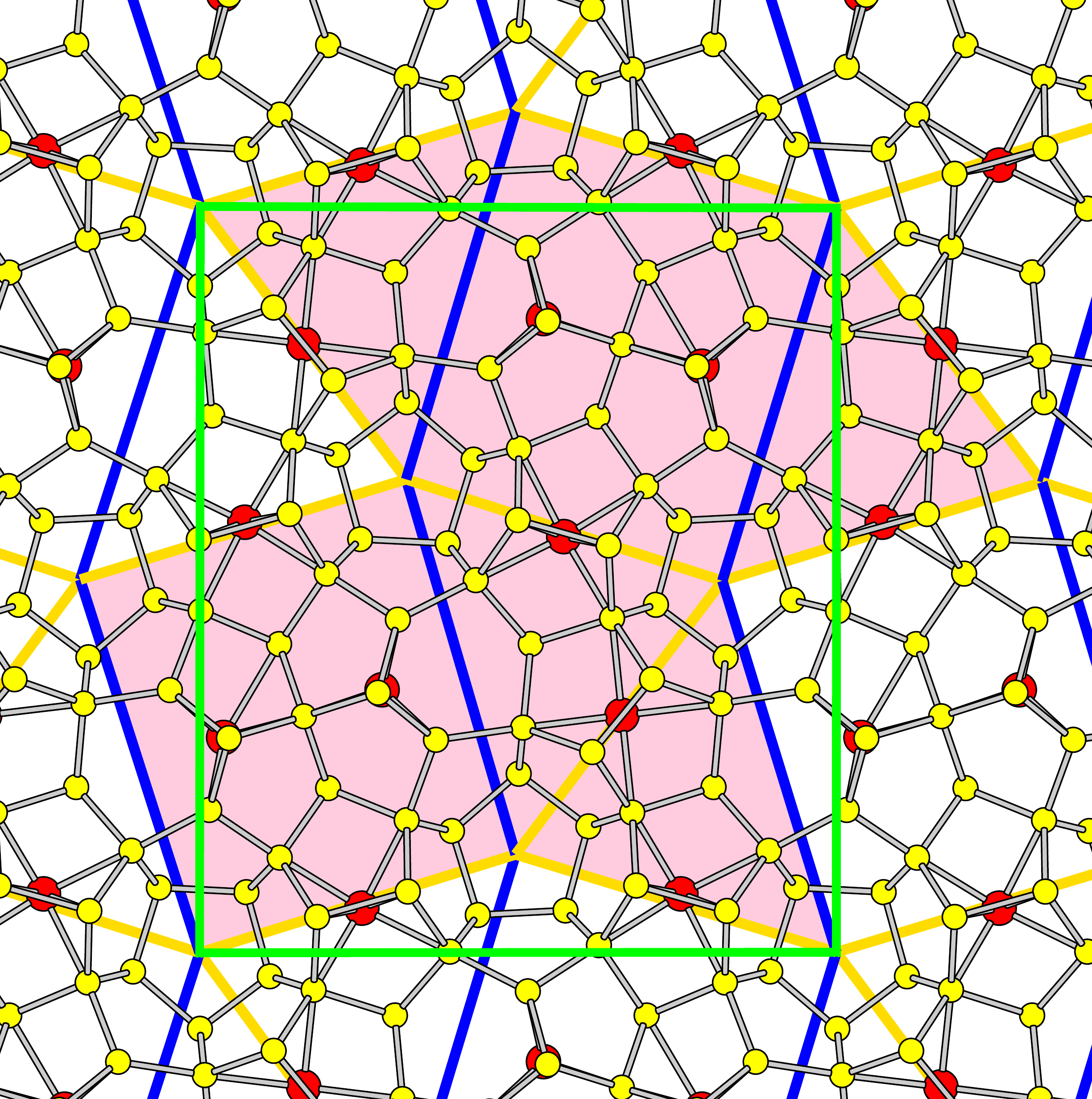}
		\caption{
			4T$'_h$+2R$'_h$ ({\em left}) and  
			~4T$_h$+2R$_h$ ({\em right})  terminations of the R$_2$T$_4$ clathrate structure. Adatoms are shown
			by red circles, a green rectangle outlines the periodic boundary in $ab$ plane.}   
		\label{penta-holes}
	\end{figure}

	The relaxed configurations of all possible surface terminations of the	R$_2$T$_4$ approximant mixing together R and T tiles are shown in 	Figs.~\ref{penta-caps} and~\ref{penta-holes}. The bulk R$_2$T$_4$ structure is shown in Fig.~S8 of SM~\cite{Supp} (VASP CONTCAR file is included in the SM named as ``CONTCAR1"~\cite{Supp}). It has a space group $Pbam$ (\#55) with 148 atoms in the unit cell. The optimized lattice parameters are $a_{ortho}$= 2.398 nm, $b_{ortho}$= 2.055~nm and	$c_{ortho}$= 1.244~nm, 
	~  $c_{ortho}$ being the stacking direction.	The surface terminations as defined in Fig.~\ref{fig:tiles}~ combine	two properties: ``penta-cap'' or ``hole'' termination of the vertex column of dodecahedra, and primed/non-primed tile flavor (pink/light blue colors). From the edge-marking rules (green symbols in	Fig.~\ref{fig:tiles}), all surface tiles in R$_2$T$_4$ tiling must have the same coloring -- this is a non-generic property of a small periodic approximant -- see representative example of large tiling from Fig.~\ref{fig:structure} in which pink/blueish tiles mix in a single layer.
	~This leaves us with four terminations, that we	implemented in three slab structures: both ``penta-cap'' terminations	in Fig.~\ref{penta-caps} are bottom and top surface of the same 262-atom slab (made of 4$F$+3$P$ type layers). 
~The primed ``penta-hole'' termination occurring on both surfaces of a	292--atom slab symmetrically, are made of 3$F$+4$P$ type layers	(Fig.~\ref{penta-holes} {\em left}, blue tiling), and finally
~the non-primed ``penta-hole'' reconstruction terminates a thinner	(2$F$+3$P$, 220 atoms) slab structure (Fig.~\ref{penta-holes} {\em right}, pink tiling).
~It may be noted that each of the four surfaces of the R$_2$T$_4$ approximant in  Figs.~\ref{penta-caps},\ref{penta-holes}  can be combined with numerical results from Table II following R/T labels given in the figure captions: the surface in Fig.~\ref{penta-caps} (left) has $\gamma$= 26 meV/\AA$^2$, and all of the three other surfaces have $\gamma$= 22-23 meV/\AA$^2$. 

	Our final remark is that the R$_2$T$_4$ tiling nearly exactly matches	the lattice parameters and atomic motifs in W-Al-Ni-Co approximant, serving	as an important hint for nucleation of the decagonal clathrate on 
	$d$-Al-Ni-Co, as discussed in the subsection~\ref{section:nucleation}.	Combining $\rho_\alpha$ for R and T tiles from Table~\ref{tab:tiles} 	with R$_2$T$_4$ tile counts and areas, we obtain coverings of $\sim$0.4
	ML$_\alpha$ for $F$ type layer and 1.09 for $P$ type	layer. Fig.~\ref{0.2MLSTM_nucl}(b) shows that the initial stage of	the decagonal clathrate formation is $F$ type layer.

\subsection{Nucleation of Sn on \anc}
\label {section:nucleation}
		\begin{figure}[tbh]
		\centering
		\includegraphics[width=145mm,keepaspectratio]{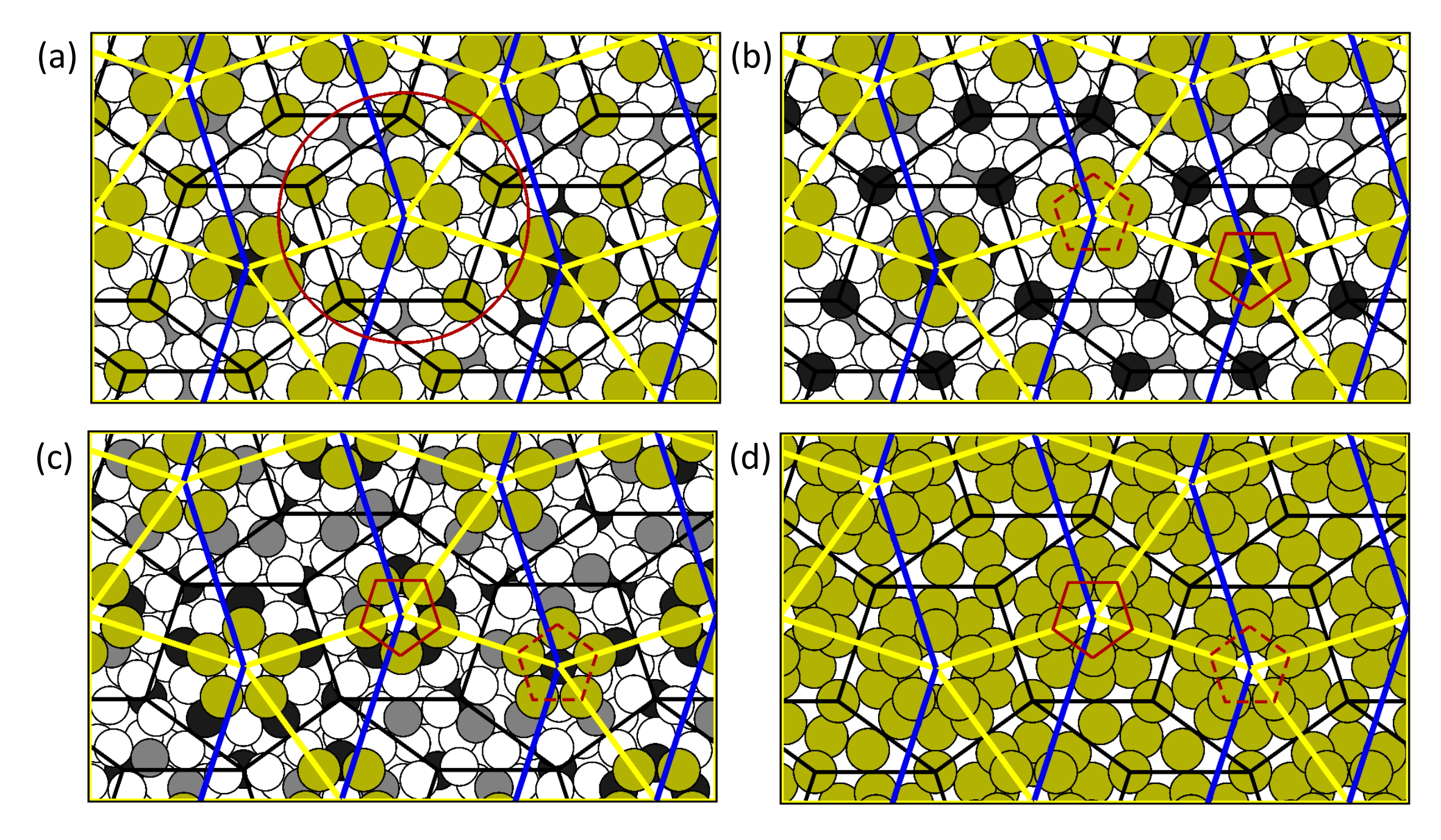}
		\caption{Nucleation of Sn on the \anc\, surface represented by decagonal W-Al-Ni-Co approximant. Here, Al atoms are represented by open circles; Co by black,	 Ni by gray  and Sn atoms by dark yellow filled circles:  (a) Sn white flower (enclosed by a red circle) on the \textit{B}-surface of the W-phase.  Sn pentagonal arrangement (solid and dashed red pentagons) on the (b) \textit{B}-surface and (c) \textit{A}-surface.  (d)  The atomic structure of the quasiperiodic Sn clathrate. The configuration of the Sn pentagon (dashed and solid red pentagon) is a part of the clathrate structure. Tilings are indicated by lines as Penrose (black) and rectangle triangle (R-T) tiling (blue and yellow).} 	
		\label{0.2MLSTM_nucl}
	\end{figure}


	Here, we focus on the question of how the decagonal Sn clathrate nucleates on the 10f \anc\, surface. As mentioned earlier, the bulk structure of  \anc\, has been described by a decagonal W-Al-Ni-Co approximant phase that can be cleaved at the flat \textit{A} surface or at the puckered \textit{B} surface~\cite{Krajci06,Sugiyama02}. So, it can be expected that the local arrangement of atoms on the \anc\, surface is similar to that in  W-Al-Ni-Co. We discuss henceforth the nucleation sites of Sn on  W-Al-Ni-Co, and how that is compatible with formation of the Sn clathrate layer.  The arrangement of Ni and Co \textit{i.e.} the transition metal (TM) atoms on the \textit{B}-surface  of W-Al-Ni-Co [see Fig.~1(b) of Ref.~\onlinecite{Krajci06}]  can be described by the P1 tiling. On the Al-Ni-Co surface, the adsorbed Sn atoms prefer to bind with the TM atoms. The preferential adsorption of Sn atoms around surface TM atoms with formation of  Sn white flowers (SnWFs) at the early stages of the Sn deposition has been  observed  on  \alpm~\cite{Singh_prr20}.  
	 		
	 On the \textit{B} surface, there are two  kinds of the preferred adsorption sites for Sn atoms. The first one is similar as that on the \alpm\ surface.  The Sn atoms can be adsorbed around the TM atom  centering a pentagon of Al atoms. The second kind is a small pentagon of TM atoms, where the Sn atoms are adsorbed in their bridge positions, compared to the clean \textit{B} surface [see Fig.~1(b) of Ref.~\onlinecite{Krajci06}]. Sn atoms can be adsorbed in the vertices of the P1 tiling and together with the small Sn pentagon in the center they could create the SnWF configuration [Fig.~\ref{0.2MLSTM_nucl}(a)], which is similar as  on the \alpm~ surface. 	On the other hand, Fig.~\ref{0.2MLSTM_nucl}(b) shows that on	 the \textit{B}-surface,  small Sn pentagons  can be adsorbed at the centers of the pentagons of the P1 tiling  (black lines). These  pentagons can	play the role of nucleation centers for growth of the Sn clathrate	layer.  The pentagon of Sn atoms can be adsorbed also in bridge positions of small Al pentagons (\textit{e.g.}, in the 	center of the figure) but such adsorption site is not preferred and not so stable as at the TM atoms. 
	
	 If the surface in the center of the P1 pentagon is occupied by a slightly protruding Co atom [right in Fig.~\ref{0.2MLSTM_nucl}(b)], the small Sn$_5$ cluster (traced by red down pentagon) formed around this central atom is stable. The calculated average binding energy ($E_b$) of Sn adatoms is $E_b$(Sn$_5$)= -0.501 eV/atom.  If the center of the P1 pentagon is occupied by Al atoms [in the center of Fig.~\ref{0.2MLSTM_nucl}(b)] the position of the Sn$_5$ adatoms (traced by dashed red up pentagon)  is metastable with  $E_b$(Sn$_5$)= -0.277 eV/atom. However, it was observed that in this case the outer Sn atoms, which occupy the vertices of the P1 pentagons [marked by a red circle in Fig.~\ref{0.2MLSTM_nucl}(a)] can significantly stabilize the central Sn$_5$ cluster. Binding of outer 5 Sn atoms with the central  Sn$_5$ cluster forms the white flower (WF) configuration of 10 atoms (known also as the starfish cluster~\cite{Ledieu2009}). 
  The WF configurations are  frequently observed at deposition of adatoms on quasicrystalline surfaces~\cite{Ledieu2009, Smerdon2008, Maniraj2014}. The calculated binding energy of the WF cluster around the  metastable Sn$_5$ is $E_b$(Sn$_{10}$)= -0.396 eV/atom. The WF cluster around the most  stable Sn$_5$ cluster [right in Fig.~\ref{0.2MLSTM_nucl}(b)] reduces the average binding energy, to $E_b$(Sn$_{10}$)= -0.426 eV/atom. The WF configurations help to create the Sn$_5$ clusters in the centers of the P1 tiles and can thus support the nucleation and further growth of the clathrate structure.
	 
	 \begin{figure}[t!]
	 	\centering
	 	\includegraphics[width=145mm,keepaspectratio]{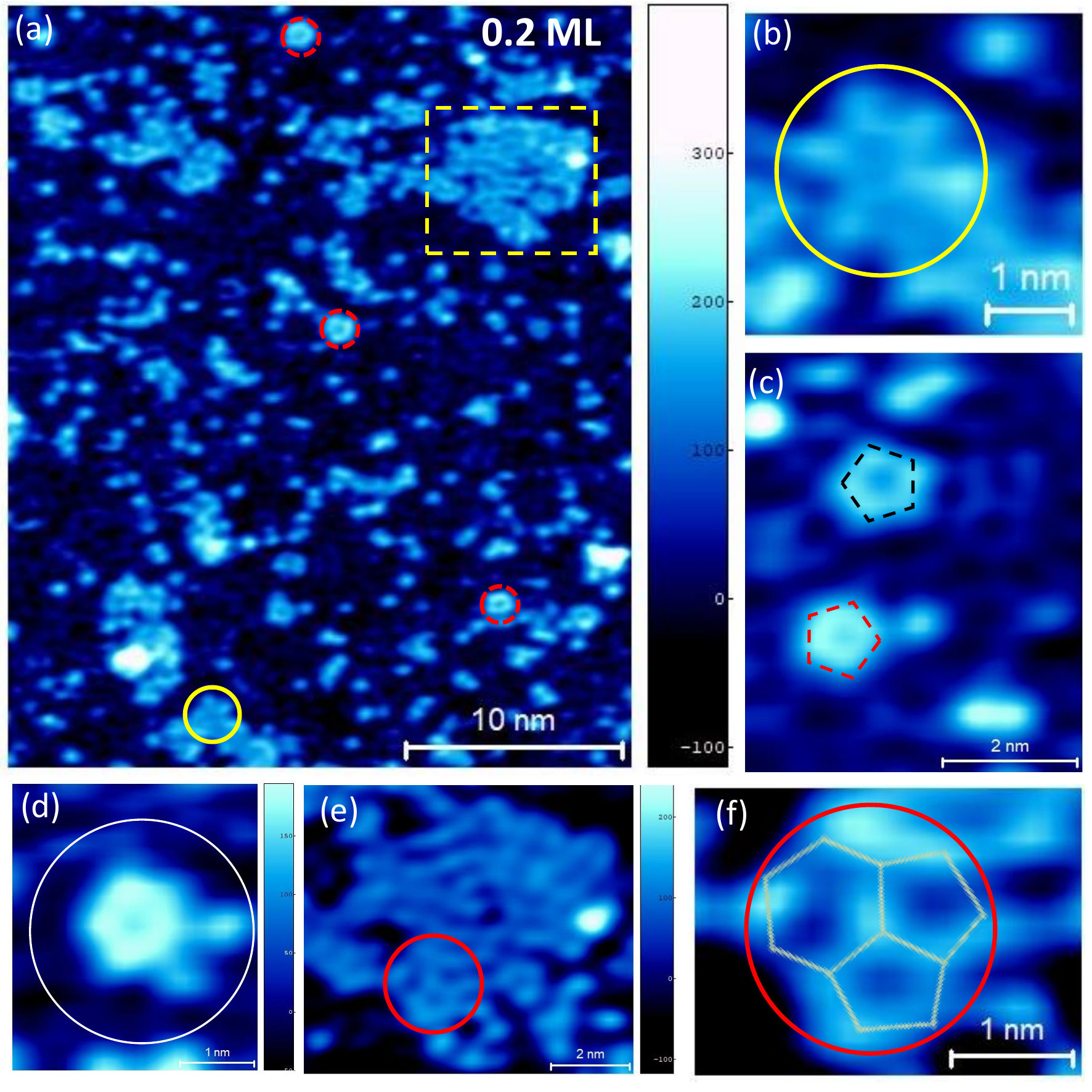}
	 	\caption{ (a) STM topography image of 0.2 ML Sn/$d$-Al-Ni-Co with $I_T$= 0.5 nA, $U_T$= -1.5 V.  Sn white flower (yellow circle, zoomed in panel \textbf{b}) and pentagonal motifs (dashed red circle).  (c) Two pentagonal motifs (traced by dashed red and black lines)  are rotated by 36${^\circ}$ with respect to each other. (d) Sn atoms attached to the pentagonal motifs. (e) The region inside the dashed yellow rectangle in panel \textbf{a} shown in an expanded scale. A triplet motif comprising of two pentagons and a hexagon (red circle)  is zoomed and compared with  the R$_2$T$_4$ model  in panel \textbf{f}.}
	 	\label{0.2MLSTM_RT}
	 \end{figure}
 
On the \textit{A} surface of W-Al-Ni-Co,  the R-T tiling can also describe ordering of	atoms on a larger scale but the local arrangement of atoms inside the tiles is less regular. Although the description of arrangement of the TM atoms on the \textit{A} surface is described by its own tiling, and for the current  discussion it is better to put on	the \textit{A} surface together with the R-T tiling also the same P1 tiling as on the \textit{B} surface. On the \textit{A} surface,  there is a higher content of TM atoms	compared to the \textit{B} surface. Similarly as on the \textit{B} surface, they can form small TM pentagons (\textit{e.g.} in the center of the figure) [Fig.~\ref{0.2MLSTM_nucl}(c)]. These small TM pentagons can be preferred adsorption centers for the Sn atoms adsorbed	in the bridge positions between the TM atoms. Fig.~\ref{0.2MLSTM_nucl}(c)	shows the \textit{A} surface with the Sn pentagons in vertices of the R-T tiling.	Similarly as on the \textit{B} surface, these can be the possible nucleation centers	for growth of the clathrate layer on the \textit{A} surface.  However, on this surface there are also additional TM atoms (inside the thin rhombi of P1) which can disturb the presented regular ordering of the Sn 	nucleation centers.

	Fig.~\ref{0.2MLSTM_nucl}(d) shows the Sn clathrate layer that is described by the  R-T tiling model (thick  blue and yellow lines).  The orientation of the Sn pentagons alternate with the number of the Sn clathrate layers. The present one is for one or odd number of layers.  Further, on the \textit{B} surface of the W-phase, the length of the Penrose P1 tiling is $d$(P1)= 0.758 nm~\cite{Krajci06}. From this it follows that the lengths of the blue and yellow edges of the R-T tiling [Fig.~\ref{0.2MLSTM_nucl}] are  $d$(blue)= $\tau$\,$\times$\,$d$(P1)= 1.227 
	~and  $d$(yellow)= $d$(blue)/[2\,$\times$\,sin($\pi$/5)]= 0.851
	$\times$\,$d$(blue)= 1.043 nm, 
	~respectively. The length of edges of the R-T tiling on the \anc\, surface is thus 2.1\% smaller compared to the \alpm\, surface~\cite{Singh_prr20}, indicating that they are almost the same. The sizes of the R-T tilings on the W-phase and \anc\, is likely to be within 1-2\%.    It is remarkable that the lengths of the R-T edges for W-Al-Ni-Co  surface ($d$(yellow)= 1.043 nm and $d$(blue)= 1.227 nm)  are so close to those of the decagonal Sn clathrate obtained from DFT in subsection \ref{subsubsec:structure} ($a$= 1.082 nm and $b$= 1.268 nm), see also Ref.~\onlinecite{mismatch}.

	 To check the compatibility of the SnWF and the  pentagon with the	 clathrate structure, it is important to compare the configuration of Sn adatoms forming the SnWF cluster and the pentagon  on the  \textit{A}  and \textit{B} surface of W-Al-Ni-Co [Figs.~\ref{0.2MLSTM_nucl}(a-c)] 	 with the structure of the clathrate layer [Fig.~\ref{0.2MLSTM_nucl}(d)]. The clathrate structure consists of regular dodecahedral cages of Sn atoms. The dodecahedra are centered at the vertices of the rectangular-triangular (R-T) tiling. It is clear from Fig.~\ref{0.2MLSTM_nucl} that SnWF and the pentagons are the integral parts of the clathrate surface.
		 
		 In order to experimentally probe the nucleation of the Sn clathrate on \anc,  we have performed STM after 0.2 ML Sn deposition   (Fig.~\ref{0.2MLSTM_RT}). A large part of the image is  dispersed  with isolated Sn atoms that have an atomic height of 0.17$\pm$0.02 nm after averaging over 110 height profiles over different Sn adatoms. However, in some regions  nucleation of Sn pentagons is clearly observed as highlighted by dashed red circles in Fig.~\ref{0.2MLSTM_RT}(a). The average diameter of the dashed red circles that enclose these  pentagons is 1.5$\pm$0.1 nm. Furthermore, a SnWF of about 2 nm diameter [yellow circle in Figs.~\ref{0.2MLSTM_RT}(a,b)] is also observed, as discussed above for the   W-Al-Ni-Co \textit{B} surface [red circle in Fig.~\ref{0.2MLSTM_nucl}(a)].  In Fig.~\ref{0.2MLSTM_RT}(c),  two  pentagons that are 36${^\circ}$ rotated with respect to each other are traced by dashed red and black pentagons, respectively. It may be noted that such rotated pentagonal motifs are also expected from our nucleation model discussed above  for both \textit{B} and \textit{A} surfaces  [dashed and solid red pentagons in Figs.~\ref{0.2MLSTM_nucl}(b,c)].  Fig.~\ref{0.2MLSTM_RT}(d) shows how a pentagonal motif converts to a SnWF  with Sn atoms bonding at its vertices. A region where Sn motifs  coalesce to form  an island [dashed yellow rectangle in Fig.~\ref{0.2MLSTM_RT}(a)] is shown in an expanded scale in Fig.~\ref{0.2MLSTM_RT}(e). Here,  motif (red circle) formed by two pentagons and a hexagon is visible, shown zoomed  Fig.~\ref{0.2MLSTM_RT}(f). This, named as triplet motif,  is compared with and shown to be a valid part of both penta-cap and penta-hole  terminations of the R$_2$T$_4$ approximant in Fig.~\ref{motifs_theory}. Thus, the indication of formation of the clathrate structure is obtained even from small Sn islands at initial stage of the growth.
		
\subsection{R$_2$T$_4$ approximant model of the Sn film on Al-Ni-Co substrate }
\label{subsec:Snonsubstrate}

W-phase approximant model structure~\cite{Sugiyama02} with its
	~530 atoms/cell utilized as a model substrate for nucleation in  subsection \ref{subsec:surfmodel}, strains the resources when the computationally
	intensive \textit{ab-initio} approach is needed. At the same time, the W-phase periodic cell is {\em double superstructure}
	~ of the R$_2$T$_4$ approximant cell (the horizontal axis of the W-phase  periodic cell in Fig.~\ref{0.2MLSTM_nucl}  is doubled along the horizontal axis relative to the R$_2$T$_4$ cell in Figs.~\ref{penta-caps}, \ref{penta-holes}). So, 
	possibly a hypothetical R$_2$T$_4$ Al-Ni-Co approximant preserves	equally fair relationship with the $d$-phase while requiring only	$\sim$10\% of the resources needed to study composite Sn/Al-Ni-Co
	slab within our DFT setup.  The R$_2$T$_4$ Al-Ni-Co should not	be confused with the R$_2$T$_4$ clathrate approximant because	although they share the tiling geometry, the atomic motifs behind		this coincidence are very different. 

\begin{figure}
	\centerline{\includegraphics[width=16cm]{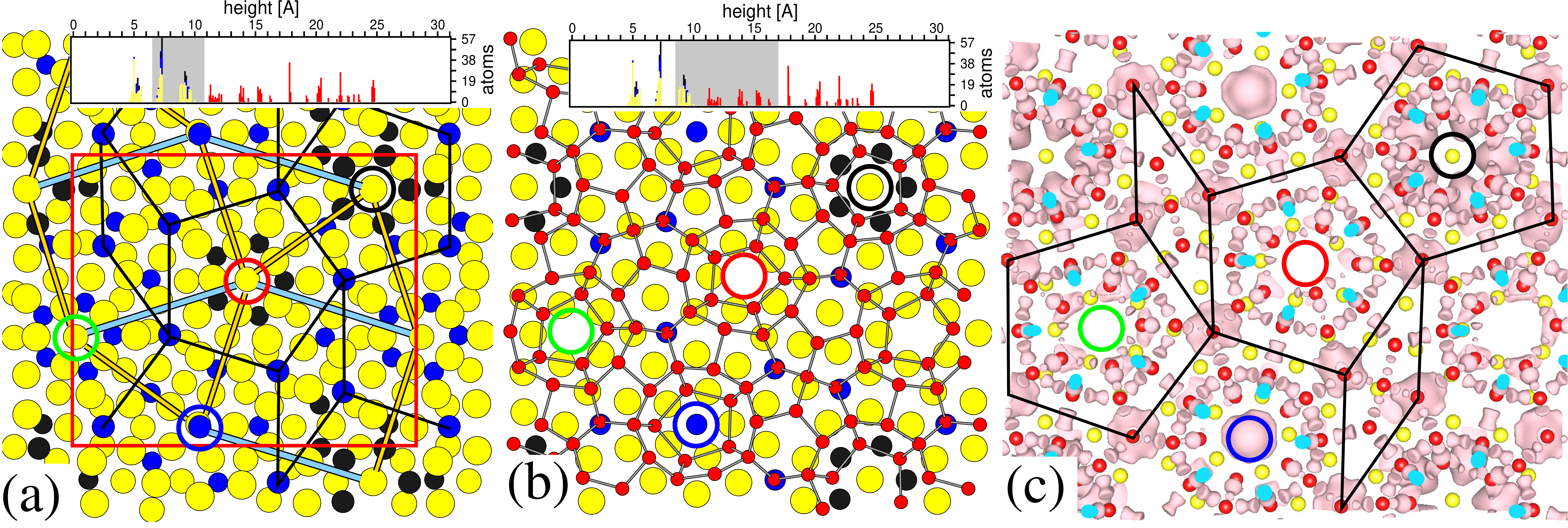}}
	\caption{\label{composite} R$_2$T$_4$ approximant		Sn/Al-Ni-Co slab structure after DFT relaxation:
		~ (a) Al-Ni-Co	``top'' bilayer near the interface; (b) slice including ``topmost'' Al-Ni-Co layer with ``bottom'' $FP$ layers of the Sn film. Height	histograms inserted at the top of panels indicate displayed slice selection by grayed background.  Atoms are represented by colored circles: Al (yellow), Co (blue), Ni (black), Sn (red), the same for histogram bars representing counts of the atoms.	(c) Pink isosurfaces of the valence charge density for the	slice shown in the panel \textbf{b}. Cyan areas indicate section opening view {\em inside} the	isosurface.}
\end{figure}

 We carried out tempering simulations for the Al-Ni-Co system in	the R$_2$T$_4$ cell in the spirit of the Al-Cu-Fe simulated	annealing leading to spontaneous formation of icosahedral	quasicrystal~\cite{MW-alcufe2020}, using DFT-fitted empirical	oscillating potentials (EOP)~\cite{MH-2012} for	Al-Ni-Co. Fig.~\ref{composite}(a) displays $R_2T_4$-Al-Ni-Co
	bilayer at the $B$--type surface analogous to its W-phase
	counterpart in Fig.~\ref{0.2MLSTM_nucl}(b). Details of the	preparation of the R$_2$T$_4$-Al-Ni-Co model are gathered as a note in the SM~\cite{Supp}, and the final DFT-relaxed structure is provided in the SM as ``CONTCAR2"\cite{Supp}.  

Although atomic structure of $d$-Al-Ni-Co and Sn clathrate obey	completely different electronic and coordination rules, their	interface is perfectly coherent following the R-T tiling geometry, as indicated in Fig.~\ref{0.2MLSTM_nucl}. This allows for a straightforward approach	to the Sn/Al-Ni-Co $R_2T_4$ approximant composite model	construction: we merge a 3--layer thick Al-Ni-Co slab (two	Al--rich ``puckered'' layers with ``flat'' pseudo-mirror layer in	between) with a five--layer 1.2~nm thick Sn slab (three F--type and	two P--type layers). The slabs were positioned within plausible distance from each other, with shortest initial pair distances of
$\sim$0.27~nm between Sn and Al/Co/Ni atoms. The open-surface side of the Sn slab was reconstructed following the primed-penta-cap recipe illustrated in Fig.~\ref{penta-caps} {\em left}, and 1.2~nm thick vacuum slab was inserted between the Sn surface and the ``bottom'' of the periodic image of Al-Ni-Co slab.  The resulting composite slab had Sn$_{180}$/Al$_{150}$Co$_{32}$Ni$_{18}$ content	in the $a_s$= 2.34 nm, $b_s$= 1.98 nm and $c_s$= 3.09~nm	orthorhombic cell (the $c_s$ axis is parallel to the pseudo-10-fold	axis), and its ``bottom'' $PF$--layers {\em relaxed} structure is	presented in Fig.~\ref{composite}(b). The final optimized structure can be found in ``CONTCAR3"~\cite{Supp}. 
	
 		The structural relaxation was executed under our DFT setup, using 4		and finally 16 $k$-points mesh, but with the rectangular cell fixed at		bulk Al-Ni-Co-$R_2T_4$ optimal $a_s\times b_s$ 
	~ size.  During the relaxation, the structure settled quickly into a local minimum with negligible residual forces and rather small -3 kBar pressure. In	order to estimate strength of the binding between Sn/Al-Ni-Co slabs, we
	separated Sn and Al-Ni-Co sub--slabs by another 1.2~nm --wide vacuum gap, and re-relaxed the structure (holding the cell fixed). We find the ``work of separation'' (or {\em	adhesion}) defined as $W_{sep}= \Delta E/A\sim$60.4 meV/\AA$^2$ (area $A= a_sb_s$ and $\Delta E$ between composite slab and	vacuum--separated Sn and Al-Ni-Co slabs in eV {\em per cell}) to be
	nearly identical to the $W_{sep}\sim$64.4 meV/\AA$^2$ 
	~ of the pure	clathrate II structure normal to its 3--fold cubic direction.  This shows that the strength of the attachment between the Sn-clathrate film	and the Al-Ni-Co substrate is close to the self- separation of the clathrate	itself. 

The gross picture offered by charge--density isosurface taken at $\rho_{el}/\rho_{el}^{max}$=0.02 ($\rho_{el}$ denotes the valence	electron density) and displayed in Fig.~\ref{composite}(c)	(slice widths as in panel \textbf{b}) is as follows: the charge mainly accumulates around transition metal (TM) atoms (that are hidden	inside the isosurfaces), with lobes of the surfaces pointing toward the Al or Sn atoms, and indicating a charge-transfer from the Al/Sn	atoms.  Second, with the exception of the ``bottom''--most Sn atoms,	characteristic dumbbell-like charge pockets accumulate around the	Sn--Sn bond mid-points, indicating clear $sp^3$ character of these	electronic states.  Side--view of the isosurfaces (not shown)	reveals that the crucial interface bonding occurs via strong	covalent 0.27~nm bonds normal to the interface between Sn and Co atoms, located at the vertices of the $P1$ tiling (black lines in	the panel \textbf{(c)}): these are the well-known $WF$--cluster tips introduced in the previous section.

To summarize, DFT study of the Sn/Al-Ni-Co interface based	on the $R_2T_4$ tiling geometry confirms that (i) there are	essentially two types of Sn atoms in the first ML: isolated Sn at	the vertices of the $P1$ tiling above Co atoms bonding to them by	strong covalent bonds; and Sn-pentagons at the vertices of the R-T	tiling, occurring above several distinct pentagonal Al-Ni-Co motifs,	supporting the findings of the previous section; (ii) above this	first ML, Sn builds up stable $sp^3$-bonded clathrate structure.

	\subsection{Motifs from STM compared to the relaxed R$_2$T$_4$ structural model}
\label{subsec:theoryandstm}

\begin{figure*}
	\includegraphics[width=160mm,keepaspectratio]{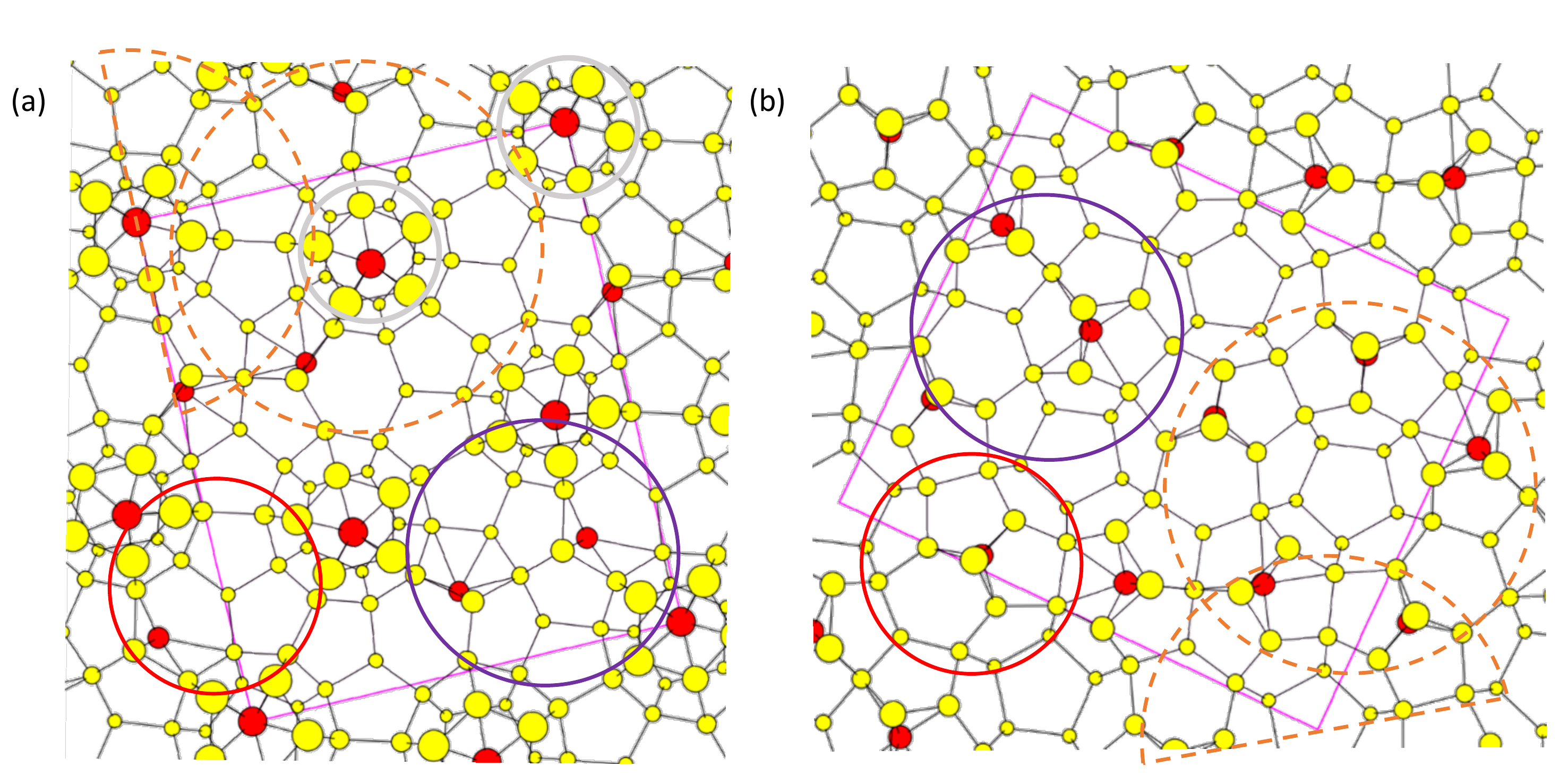}
	\caption{ Energy-optimized relaxed surface of the R$_2$T$_4$ clathrate approximant   with (a) ``penta-caps'' reconstruction and (b)  ``penta-hole'' reconstruction, the pink rectangle shows the unit cell. The motifs are highlighted by circles: wheel (dashed orange),  polygon assembly (violet), triplet (red),  pentagon (gray) and crown (dashed orange   half-circle). The sizes of the yellow filled circles representing the Sn  atoms scale with their vertical height, red filled circles mark the Sn adatom positions.}
		\label{motifs_theory}
\end{figure*}

\begin{figure*}[t]
	\includegraphics[width=160mm,keepaspectratio]{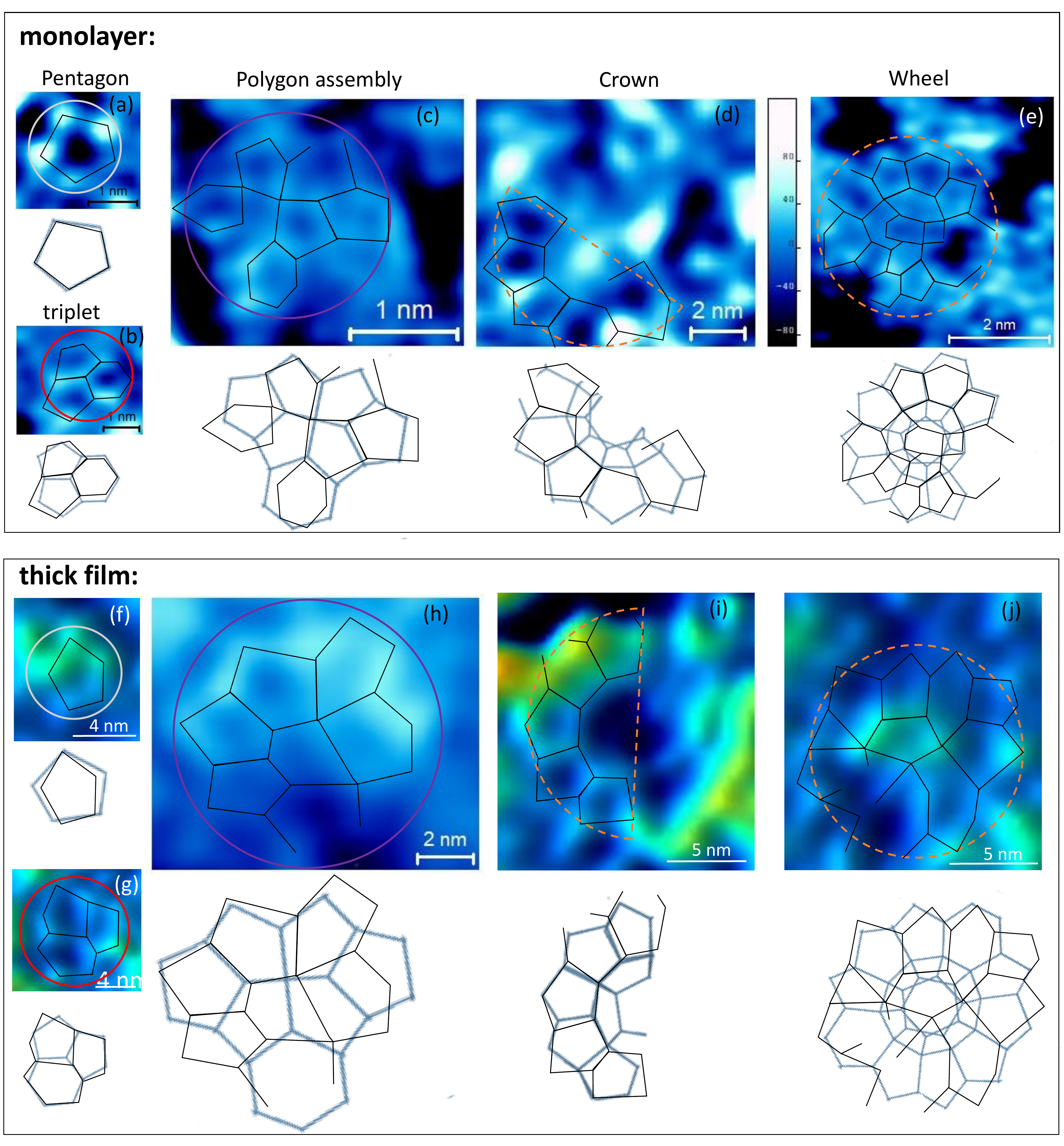}
	\caption{The  different types of motifs  (indicated at the top of each figure column) such as pentagon (gray circle),  triplet (red circle), polygon  assembly (violet circle), crown (dashed orange half-circle), and wheel  (dashed orange circle)  observed from STM for  (a-e) the  monolayer and (f-j) the 10 nm thick film. Each motif is traced by black lines [some untraced images are provided in Figs.~S9(e-h) for comparison] and  juxtaposed on the relaxed surface of the R$_2$T$_4$ clathrate approximant (light blue lines) in the same scale below each STM image. The R$_2$T$_4$  approximant was inflated by $\tau^2$ and $\tau^4$$\chi$  for the monolayer and the thick film, respectively. } 
	\label{motif_comparison}
\end{figure*}

\begin{figure*}
	\includegraphics[width=100mm,keepaspectratio]{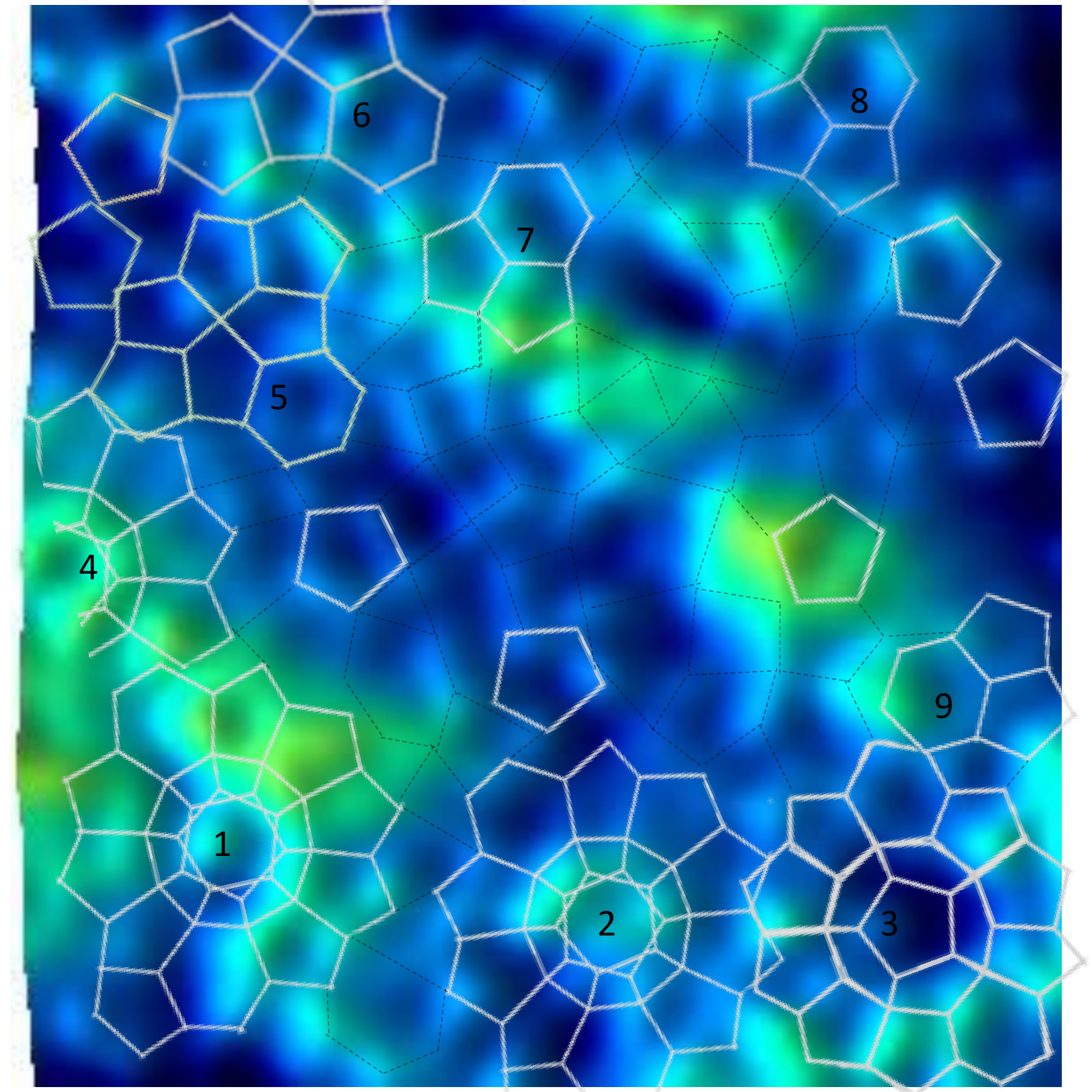}
	\caption{A portion of the STM image of the 10 nm thick Sn film  [enclosed by white dashed rectangle in Fig.~\ref{thickSTM_LT}(d)]  compared with motifs such as wheel (numbered as \#1- \#3), crown (\#4), polygon assembly (\#5- \#6), and triplet (\#7- \#9)  after inflation (white lines).  The  regions that are not in agreement with  the R$_2$T$_4$ approximant model are traced by black dashed lines.} 
	\label{motif_conta}
\end{figure*}

In this section, we define the  different types of motifs, each highlighted  by similar line-type in the STM images in Figs.~\ref{1ML_STM_RT}, \ref{leed_stm_intrm}, \ref{thickSTM_LT} and \ref{0.2MLSTM_RT})  based on  the energy-optimized relaxed surface of the R$_2$T$_4$ clathrate approximant in Fig.~\ref{motifs_theory}. Thereafter,  a  comparison is provided  between STM and theory 
~in Figs.~\ref{motif_comparison}, \ref{motif_conta}. 

The \textit{wheel} motif enclosed by  dashed orange circle in Fig.~\ref{motifs_theory} is a nearly decagonal congregation of 10 polygons (8 pentagons and two hexagons) with common sides. Sn atoms decorate their vertices. Incomplete wheels with more than 5 polygons are also referred to as the wheel motif. Also note that the polygons  are not regular, the length of their sides have a mean value of  0.3~nm with standard deviation of  $\pm$0.01 nm [Fig.~\ref{motifs_theory}].  A perfectly decagonal wheel with ten pentagons does not occur in the R$_2$T$_4$ approximant, but this is a feasible object if a bigger approximant is considered. It simply corresponds to a particular tiling pattern, namely five triangles forming pentagons with blue tiling edges.  The  wheel motifs are found on both the “penta-cap” and “penta-hole”  surfaces, as shown in  Fig.~\ref{motifs_theory}(a,b), respectively. In the former, the decagon is capped  with the possible occurrence of  adatom, which would give a bright contrast at the center in the STM image. In the latter, the decagon is not capped,  resulting in a relatively darker center of the wheel motif. The \textit{crown} motif is  a part of the wheel motif comprising of $\leq$5 polygons highlighted by a dashed orange half-circle in  Fig.~\ref{motifs_theory}.    The \textit{polygon assembly} with variable number of polygons  comprises of the overlapping regions of  two adjacent wheel motifs and consist of pentagons with/without a hexagon  (violet circles).  The \textit{triplet} motif  is a smaller version of the polygon assembly, comprising of two pentagons and one hexagon (red circles).  The \textit{pentagon}  motif has its  vertices decorated by the five top most  Sn atoms  in the center region of the “penta-cap” wheel , as shown by a gray circle in Fig.~\ref{motifs_theory}(a). 

The above defined motifs are compared with the STM images in Fig.~\ref{motif_comparison} for two different thicknesses. It is  interesting to observe that  there is an inflation between the motifs of the monolayer [Fig.~\ref{motif_comparison}(a-e)] and the 10 nm film [Fig.~\ref{motif_comparison}(f-j)], which  is estimated to be \AC$\tau^2\chi$.  For example, the average  diameter of the wheel motif is  4.5$\pm$0.5 nm for monolayer, 8.8$\pm$1 nm for 0.9 nm and 14.5$\pm$1~nm for the 10 nm film.  This shows  inflation by a factor of 3.2$\pm$0.4 between monolayer and the 10 nm film, which  is  close to  $\tau^2$$\chi$ (= 3.08).  Similarly,  inflation by a factor of 1.65$\pm$0.2 between the 0.9 nm and the 10 nm film   is  close to   $\tau$ (= 1.618).   This is consistent with the finding that the theoretical model  requires an inflation of $\tau^2$ for good agreement with the sizes the  monolayer motifs, whereas it is $\tau^3$$\chi$ for the 0.9 nm film and $\tau^4$$\chi$ for the 10 nm film.

Inflation is a well known property of  the quasiperiodic systems due to their self-similar nature  and has been observed in both ternary and binary quasicrystals~\cite{Krajci_08, Tsai_15, Takakura_nm07, Madison_pss14,Singh_prr20}.  A well-known example is $i$-Al-Pd-Mn, where the fundamental inter cluster linkage of 7.75\,\AA\, is $\tau$ ($\tau^3$) inflated along the twofold (fivefold) direction.  $\tau^3$ inflation has also been reported in a binary quasicrystal Yb-Cd with the formation of a cluster of clusters~\cite{Takakura_nm07}.  Inflation could be caused with respect to the theoretical model by the formation of self-similar structures  with the inclusion of extra atoms such that the local bonding characteristics remain similar.  
  In our previous work, we showed that the STM image from 3-4 nm thick Sn film on $i$-Al-Pd-Mn shows $\tau^3$ inflation with respect to the theoretical clathrate model~\cite{Singh_prr20}. Here, from Fig.~\ref{motif_comparison}, we find satisfactory agreement in the sizes of the STM motifs  and the  R$_2$T$_4$ model  at a fixed inflated length scale for all the motifs at a particular thickness. This  is an indication of the propagation of  quasiperiodic ordering  in the film.

The pentagon motif in Figs.~\ref{motif_comparison}(a,f)  is the smallest of all the motifs and is traced by black lines that join the bright vertices. It is also the most abundantly observed motif, for example, see  Fig.~\ref{thickSTM_LT}(d).  The tracing of the triplet motif  by black lines revealing its internal structure is shown in Figs.~\ref{motif_comparison}(b,g). 
	~The tracings of these  motifs are juxtaposed on the R$_2$T$_4$ model below each image in same length scale, and the agreement is rather good for both the thicknesses vindicating the clathrate model. Note that the triplet motif is also observed  for 0.2 ML Sn deposition in Fig.~\ref{0.2MLSTM_RT}(f), and the agreement with the R$_2$T$_4$ model is satisfactory. 

The polygon assembly motif is also observed in both the monolayer as well as the thick film, as shown in Figs.~\ref{motif_comparison}(c,h), respectively. 
	~The crown motif in Fig.~\ref{motif_comparison}(d) with bright center  for the monolayer in contrast to a  dark centered one in   Fig.~\ref{motif_comparison}(g)  for the 10 nm film shows existence of both ``penta-cap" and ``penta-hole" type surfaces. A crown motif  on the 0.9 nm Sn film is shown in Fig.~S6(a) of SM~\cite{Supp}. The wheel motifs are shown in Figs.~\ref{motif_comparison}(e,j), as well as in Figs.~S4(b) and Fig.~S6(b) of SM~\cite{Supp}, the latter being for  the 0.9 nm thickness film.
When juxtaposed on the R$_2$T$_4$ model, the tracings of these bigger motifs (polygon assembly, crown, and wheel) reveal their size similarity after inflation.  Moreover,  the ratio of the sizes of the  pentagon and the wheel motifs is in good agreement between STM and theoretical model.  However, in contrast to smaller motifs such as the pentagon and triplet, the underlying structure of the larger motifs is only partially consistent with the clathrate model. Nevertheless, this partial agreement is also a significant result given that  stochastic processes and competing disorder  would have a greater influence on the bigger motifs.  Moreover, the puckering and roughness of the films could also play a role. This is also revealed in the theoretical model, where the atoms  deviate considerably from the $x$-$y$ plane:  Fig.~\ref{motifs_theory} show that the average height of  the Sn atoms is  $z$= 0.25$\pm$0.12 nm, where the standard deviation provides an  estimate of the puckering. 
	~In the R$_2$T$_4$ model, two Sn atoms, for instance, forming a side of the pentagon in the wheel motif and pointing in the radial direction, are often at a greater height than the remaining three atoms. In STM, this side would look brighter, but other sides containing Sn atoms at a lower height may appear darker. Thus, the STM pictures may not depict entire and regular polygons. The comparatively modest size of the R$_2$T$_4$ approximant may also contribute to the discrepancy between STM and the model; the internal structure of a larger approximant may result in a better agreement.

In Fig.~\ref{motif_conta}, the different motifs (numbered \#1- \#9) of R$_2$T$_4$ approximant clathrate model are overlaid on (white lines) the STM image of the Sn thick film after allowing for rotation (motifs \#2- \#9 need to be rotated  by  different angles
~with respect to motif  \#1). 
~In this way, we find that a region  of  800$\pm$50 nm$^2$ is in good agreement with our model.  Requirement of  rotation can be  related to the  
~the formation of  ring joining the spots in the LEED pattern [Fig.~\ref{thickLEED_LT}(a,b)]. The regions that do not follow the R$_2$T$_4$ model are traced by black dashed lines in Fig.~\ref{motif_conta}.  Such regions could exist because of competing disordered structures since  for the thick Sn films with  clathrate multilayers  (that has not been calculated by DFT),  the advantage of interfacial compatibility would be reduced.

Last but not least, the Sn motifs on \alpm\, that reflect the clathrate structure~\cite{Singh_prr20} and those on \anc\, exhibit a considerable degree of similarity. Figure S9 depicts this for the pentagon, triplet, crown, and wheel designs. The formation of the clathrate quasiperiodic structure on two distinct types of quasicrystalline surfaces implies that it is an inherent characteristic of Sn. 
			
\section{Conclusions}
 We have carried out a combined experimental and theoretical study 
~to probe the possible occurrence of  quasiperiodicity in  Sn films grown on \anc\, up to an average thickness of  10 nm. Decagonal spots in the LEED pattern, characteristic motifs  and  the FFT of the STM topography images establish decagonal quasiperiodicity in the  Sn thin film ($<$1 nm). For the thicker films up to  the largest thickness of 10 nm where the effective potential of the substrate   is  negligible~\cite{Ye15,Lang70}, the motifs, weak LEED and FFT are also observed, showing that partial decagonal structural correlations are maintained  in spite of  the competing disorder. 
~ The clathrate atomic positions  of  the DFT-relaxed decagonal clathrate structure reveal remarkable inflation relationship with  STM.  The internal structure of the smaller motifs such as pentagon and triplet show good agreement  with   our DFT based relaxed R$_2$T$_4$ approximant clathrate structural model, while for the larger motifs the agreement is partial.  This model is the most-likely  hypothesis for the Sn structure because (i)~compatibility of the clathrate structure with the substrate resulting from very similar edge lengths (within 1\%) of the R-T tiling of the R$_2$T$_4$-approximant  and the \textbf{B} surface of W-Al-Ni-Co approximant for the \anc\ surface,  (ii)~the ground states of the thicker slabs are $sp^3$-bonded, the metallic Sn structures are only stabilized by vibrational entropy at elevated temperatures;  (iii)~the clathrate has significantly lower surface energy than the $\alpha$-Sn, hence for sufficiently thin films clathrate structure has lower energy, and (iv)~the  “work of separation” or “adhesion” of Sn/$d$-Al-Ni-Co  turns out to be comparable to the same quantity calculated for the clathrate itself  proving that  the clathrate Sn effectively binds to the $d$-Al-Ni-Co surface.  

 \section{Acknowledgments}
M.M. and M.K. are thankful for the support from the Slovak Grant Agency  VEGA (No. 2/0144/21) and APVV (No. 15-0621, No. 19-0369). Parts of the calculations were performed in the Computing Center of the Slovak Academy of Sciences using the supercomputing infrastructure acquired under Projects ITMS 26230120002 and 26210120002. R. Batabyal is specially thanked for important suggestions and a careful reading of the manuscript.  M. Balal, P. Sadhukhan, and S. Barman are thanked for support.   The   growth of $d$-Al-Ni-Co single grain substrate  done at Ames laboratory  was supported by the U.S. Department of Energy, Office of Basic Energy Science, Division of Materials Sciences and Engineering.  Ames Laboratory is operated for the U.S. Department of Energy by Iowa State University under Contract No. DE-AC02-07CH11358.
 \vskip 10mm
 \noindent e-mail addresses: $^{\dagger}$Marek.Mihalkovic@savba.sk, 
 $^{**}$Marian.Krajci@savba.sk, $^{\#}$barmansr@gmail.com


\begin{thebibliography}{}
\bibitem{mismatch} For W-Al-Ni-Co, the lattice parameters optimized by DFT are by factor 0.9975 shorter than the experimental lattice. 	The fundamental $\alpha$-Sn structure yields optimized lattice parameters by 1.023 factor longer than the experiment.  The DFT lattice mismatch computed for the R$_2$T$_4$ approximants of the Sn-clathrate versus \anc\, is 1.033. Thus, our extrapolated realistic mismatch estimate is 1.033/1.023$\times$0.9975 = 1.007.
\end{thebibliography}
\end{document}